\documentclass[aps,prl,twocolumn,superscriptaddress,showpacs,floatfix,longbibliography,nofootinbib,preprintnumbers]{revtex4-2}
\usepackage[utf8]{inputenc}
\usepackage{graphicx}
\usepackage{color}  
\usepackage{amsmath}
\usepackage{dcolumn}
\usepackage[english]{babel}
\usepackage[colorlinks=true,linkcolor=blue,citecolor=blue, urlcolor=blue]{hyperref}
\usepackage[dvipsnames]{xcolor}
\usepackage{multirow}

\begin{document}

\title{Imaging two-body correlations in atomic nuclei via low- and high-energy processes}

\author{Stavros Bofos}
\affiliation{CEA, DES, IRESNE, DER, SPRC, LEPh, 13108 Saint-Paul-lez-Durance, France}

\author{Benjamin Bally}
\affiliation{Technische Universit\"at Darmstadt, Department of Physics, 64289 Darmstadt, Germany}

\author{Thomas Duguet}
\affiliation{IRFU, CEA, Universit\'e Paris-Saclay, 91191 Gif-sur-Yvette, France}

\author{Mikael Frosini}
\affiliation{CEA, DES, IRESNE, DER, SPRC, LEPh, 13108 Saint-Paul-lez-Durance, France}

\begin{abstract}
Characterizing  the correlated behavior of nucleons inside atomic nuclei constitutes a long-standing challenge, both experimentally and theoretically. It has recently been understood that two-particle correlations in the azimuthal distribution of final hadrons emitted in ultra-relativistic ultra-central ion-ion collisions can be used to quantify ground-state two-body correlations. Performing systematic {\it ab initio} nuclear structure calculations of light nuclei, we demonstrate that such an observable does provide a meaningful imaging of nuclear ground states, naturally leading to a robust interpretation of the various categories of two-nucleon correlations at play. This is at variance with the low-energy approach relying on Kumar operators whose traditional interpretation in terms of deformation parameters is shown to be inoperative. A future interesting development will consist of targeting specific three-particle correlations to isolate three-nucleon correlations in which additional nuclear structure information of interest leave their fingerprint.
\end{abstract}

\maketitle


\paragraph{Introduction.} 

The atomic nucleus is a strongly interacting finite quantum many-body system within which the correlated behavior of nucleons give rise to a rich variety of phenomena. Appearing at a similar energy scale, individual and collective nucleonic motions are intertwined. Furthermore, the nature of these collective correlations as well as their dominance over non-collective ones strongly depend on the atomic nucleus under consideration and varies quickly with proton and/or neutron numbers. These various characteristics strongly challenge an understanding based on clean experimental signatures and on a theoretical description connecting inter-nucleon interactions to emergent nuclear structure properties. 

Traditionally, the collective behavior of nucleons has been inferred through its fingerprints on low-energy observables. Most notably, subsets of excited states displaying specific energy spectrum and electromagnetic decay probabilities can be associated, through a simplistic model, with a rotor (i.e., rotational excitations) characterized by a given intrinsic deformation or with a harmonic oscillator (i.e., vibrational excitations) characterized by a given phonon frequency. Associated correlations can be quantified via the ground-state expectation value of so called Kumar operators~\cite{Kumar1972a,Poves2020a,Henderson2020a} that can, in principle, be reconstructed experimentally by combining electromagnetic transition probabilities to a set of excited states \cite{Cline1986a,Garrett2024a,Bree2014a,Ayangeakaa2019a,Henderson2019a,Rocchini2021a}.

Recently, however, ultra-relativistic ion-ion collisions at the Relativistic Heavy Ion Collider (RHIC) and the Large Hadron Collider (LHC) have shown to offer a new and complementary perspective on multi-nucleon correlations at play in nuclear ground states. The instantaneous character of high-energy collisions (of the order of $10^{-26}s$) is such that both nuclei remain in their respective ground state during the process. Following the formation and cool down of a Quark-Gluon Plasma (QGP), the azimuthal distribution of final hadrons emitted in the plane transverse to the beam axis can indeed be related, though hydrodynamic and hadronization simulations, to the spatial distribution of correlated nucleons at the time of the collision~\cite{Giacalone2018a,Giacalone2021a,Summerfield2021a,Bally2022a,Zhang2022a,Ryssens2023a,STAR2024a,Giacalone2025a,Giacalone2025b}. 

During the first steps of this endeavor~\cite{Giacalone2018a,Bally2022a,Zhang2022a,Ryssens2023a,STAR2024a}, the analyses were performed based on the model picture of the nucleus as a classical rotor, i.e., a rotating deformed rigid body~\cite{Duguet:2025qxi}. While this simple interpretation in terms of a model-dependent intrinsic deformation is qualitatively applicable to nuclei whose experimental spectrum display a clear ``rotational'' band on top of the ground state, it is not adapted to other nuclei, i.e., doubly-magic, semi-magic and so-called transitional nuclei. In spite of the key results such early analyses  delivered~\cite{Bally2022a,Zhang2022a,Ryssens2023a,Giacalone2025a,Giacalone2025b}, the situation calls for a more robust interpretation of the data based on a general quantum mechanical description of the colliding nuclei. 

Recently, a decisive step was taken in that direction for ultra-central symmetric collisions~\cite{Duguet2025a}. It was indeed possible to connect the fluctuation of the $\ell^{\text{th}}$ Fourier component of the azimuthal hadronic flow to the ground-state density-density correlation function of the colliding nuclei. The link goes through the convolution with an operator assessing the square of the $\ell^{\text{th}}$ moment of the eccentricity of the nucleons in the transverse plane of the reaction. This connection elucidates in which way (i) ultra-relativistic collisions provide a direct access to the harmonic spectrum of two-body correlations in nuclei and, conversely, (ii) how nuclear structure calculations properly accounting for many-body correlations are of critical importance to pin down fluctuations of the experimentally accessible flow coefficients $\upsilon_{\ell}$ characterizing the behavior of the QGP~\cite{Shuryak2017a,Schenke2010a,Busza2018a,Shen2020a,Ollitrault2023a}.

In this Letter, a systematic analysis of the mean-square eccentricity related to the mean-square elliptic flow and probing quadrupolar correlations is provided for all $0^+_1$ ground states of even-even nuclei with proton number $8\leq Z \leq 28$. Results are compared to those obtained with the second-order quadrupolar Kumar operator. The analysis is based on state-of-the-art {\it ab initio} nuclear structure calculations rooted into the low-energy Chiral Effective Field Theory ($\chi$EFT) of the strong interaction described via Quantum Chromodynamics (QCD). For $^{12}$C, $^{16}$O and $^{20,22}$Ne, results are also provided for the first excited $0^+_2$ state as well as for the mean-square eccentricity related to the mean-square triangular flow and reflecting octupolar correlations in nuclei. This analysis quantifies specific angular correlations among nucleons and demonstrates how they strongly differ depending on the category of nuclei under consideration. The extent by which quadrupolar and octupolar correlations at play in nuclei can be mapped onto intrinsic deformation parameters is elucidated. 


\paragraph{Nuclear structure calculations.}

Present {\it ab initio} nuclear structure calculations are based on the Projected Generator Coordinate Method (PGCM)~\cite{Frosini2022a,Frosini2022b,Giacalone:2024luz}. This is a variational approach based on the many-body ansatz
\begin{equation}
| \Psi^{J\Pi}_k \rangle \equiv \sum_{\beta_{20}, \beta_{30}} f^{J\Pi}_{k}(\beta_{20},\beta_{30}) P^{J\Pi NZ} | \Phi(\beta_{20},\beta_{30}) \rangle \, , \label{PGCMansatz}
\end{equation}
where $| \Phi(\beta_{20},\beta_{30}) \rangle$ denotes a set of axially-deformed Bogoliubov mean-field states breaking angular-momentum $(J)$, parity $(\Pi)$ and particle-number $(NZ)$ symmetries, and constrained to display the quadrupole (octupole) deformation $\beta_{20}$ ($\beta_{30}$) on average. The broken symmetries are restored via the application of the projection operator $P^{J\Pi NZ}$~\cite{Sheikh2021a,Bally2021b,RingSchuck} before mixing over the values of $\beta_{20}$ and $\beta_{30}$ such that the coefficients of the mixing $f^{J\Pi}_{k}(\beta_{20},\beta_{30})$ are determined variationally~\cite{RingSchuck}. 

The four nuclei $^{12}$C, $^{16}$O and $^{20,22}$Ne are indeed investigated based on the  full-fledged PGCM ansatz that has been shown to induce accurate azimuthal flow in $^{16}$O+$^{16}$O and $^{20}$Ne+$^{20}$Ne ultra-central collisions~\cite{Giacalone:2024luz,CMS:2025tga,ALICE:2025luc,ATLAS:2025meas}. The systematic over even-even nuclei ground-states with  $8\leq Z \leq 28$ is performed while limiting the sum in Eq.~\eqref{PGCMansatz} to the single term $| \Phi(\beta_{20},\beta_{30}) \rangle_{\text{dHFB}}$ corresponding to the variational minimum of the deformed Hartree-Fock-Bogoliubov (dHFB) mean-field equations. This corresponds to reducing the PGCM ansatz to the Projected HFB (PHFB) ansatz that essentially corresponds to a fully quantum rotor model. All calculations are performed with the EM1.8/2.0 $\chi$EFT Hamiltonian~\cite{Hebeler2011a}, which is known to deliver binding energies with an accuracy better than $2\%$ over mid-mass nuclei~\cite{Stroberg:2019bch}. 


\paragraph{Correlations from high-energy ground-state collisions.}

The one-body eccentricity operator of multipolarity $\ell>1$ and its fluctuation are first introduced as
\begin{equation}
\label{eq: 1-body eccentricity operator definition total}
    \mathcal{E}^{(1)}_{\ell} \equiv c_{\ell}^{-1} \sum_{i=1}^{A} r_{i}^\ell \, Y_\ell^{\ell}(\Omega_{i})\; ,\quad 
    \delta \mathcal{E}^{(1)}_{\ell} \equiv \mathcal{E}^{(1)}_{\ell} - \langle \mathcal{E}^{(1)}_{\ell} \rangle \; ,
\end{equation}
where the normalization constant $c_{\pm \ell}$ reported in the Supplemental Material (SM) is chosen in a way that ensures $Y_\ell^{\pm \ell}(\Omega)$ is normalized on the unit sphere
and  where the expectation value is taken with respect to the many-body state of interest $| \Psi^{J\Pi}_k \rangle$.

As shown in Ref.~\cite{Duguet2025a}, the normalized variance of the eccentricity of multipolarity $\ell$
\begin{align}
    \label{eq: 2-body eccentricity operator expectation value}
\langle \delta \epsilon^2_\ell\rangle &\equiv \frac{1}{2} \frac{\langle \delta \mathcal{E}^{(1)}_{\ell} \delta \mathcal{E}^{(1)}_{-\ell}  \rangle}{\langle   R^{(1)}_{\ell}  \rangle^2} \equiv  \frac{1}{2} \frac{\langle\mathcal{E}^{(2)}_{\ell}\rangle - |\langle \mathcal{E}^{(1)}_{\ell}\rangle|^2}{\langle   R^{(1)}_{\ell}   \rangle^2} \; ,
\end{align}
is, to first order, proportional to the statistical variance of the $\ell^{\text{th}}$ Fourier component $\upsilon_{\ell}$ of the  azimuthal flow by virtue of the hydrodynamic nature of the QGP. In Eq.~\eqref{eq: 2-body eccentricity operator expectation value}, the squared eccentricity operator is defined as $\mathcal{E}^{(2)}_{\ell} \equiv \mathcal{E}^{(1)}_{\ell} \mathcal{E}^{(1)}_{-\ell}$ whereas $R^{(1)}_{\ell} \equiv \sum_{i_1=1}^{A} (x^2_{i_1} + y^2_{i_1})^{\ell/2}$ stands for the one-body transverse radius operator. For $J^\Pi = 0^+$ states under present investigation, the mean eccentricity $\langle \mathcal{E}^{(1)}_{\ell}\rangle$ is zero such that second term in the numerator of Eq.~\eqref{eq: 2-body eccentricity operator expectation value} drops. As shown in the SM, $\mathcal{E}^{(2)}_{\ell}$ is the sum of a one-body operator $\mathcal{E}^{(2)\, 1\text{b}}_{\ell}$ and of a two-body operator $\mathcal{E}^{(2)\, 2\text{b}}_{\ell}$ such that the normalized mean-square eccentricity itself decomposes according to 
\begin{align}
    \label{eq: decompo eccentricity operator expectation value}
\langle \delta \epsilon^2_\ell\rangle &\equiv \langle \delta \epsilon^2_\ell\rangle^{1\text{b}} + \langle \delta \epsilon^2_\ell\rangle^{2\text{b}} \; .
\end{align}

As demonstrated in Ref.~\cite{Duguet2025a} and recalled in the SM, computing the ground-state one- and two-body local densities based on a classical rigid rotor (RR) model characterized by axial quadrupole ($\beta_{20}$) and octupole ($\beta_{30}$) deformation parameters delivers
\begin{equation}
\label{eq: mean squared anisotropy projected rigid rotor}
\begin{aligned}
    \langle \delta \epsilon^2_2 \rangle^{\text{RR}}_{0^+_1} = \frac{1}{A} + \frac{3}{4\pi} \beta^2_{20}\; ,\quad
    \langle \delta \epsilon^2_3 \rangle^{\text{RR}}_{0^+_1} = \frac{16}{3\pi A} + \frac{2048}{245\pi^3} \beta^2_{30}\; .
\end{aligned}
\end{equation}
The two contributions to $\langle \delta \epsilon^2_\ell \rangle^{\text{RR}}$ behave very differently: while the one-body contribution is equal to $A^{-1}$ and is independent of the deformation parameter, the two-body contribution probing genuine two-body correlations is independent of the mass $A$ but scales with the square of the intrinsic deformation parameter. Based on this model result, a quantity homogeneous to the square of a dimensionless deformation parameter is introduced according to~\cite{Blaizot2025a}
\begin{equation}
    \label{eq: Beta parameter definition1}
    \mathcal{B}^2_{\ell}(\text{HE}) \equiv \frac{4\pi}{3} \langle \delta \epsilon^2_\ell\rangle^{2\text{b}} \; .
\end{equation}


\paragraph{Correlations from low-energy spectroscopy.}

A model-independent way to quantify (ground-state) correlations and their collective character has relied on Kumar operators $\mathcal{Q}^{(n)}_{\ell}$~\cite{Kumar1972a}. For simplicity, the present discussion is limited to the second-order quadrupole operator $\mathcal{Q}^{(2)}_{2}$. Starting from the one-body electric quadrupole operator
\begin{equation}
    \label{eq: quadupole operator definition}
    Q^{(1)}_{2 \mu} \equiv \sum_{i=1}^{A} e_i  r^2_{i} Y^{\mu}_{2}(\Omega_{i})\; ,
\end{equation}
one defines
\begin{equation}
    \label{eq: Kumar 2-body operator definition}
    \begin{aligned}
    \mathcal{Q}^{(2)}_{2} &\equiv \sum_{\mu=-2}^{2} (-1)^{\mu} Q^{(1)}_{2 \mu} Q^{(1)}_{2 -\mu} \; .
    \end{aligned}
\end{equation}
This operator entertains a close proximity with $\mathcal{E}^{(2)}_{2}$, except that the former involves only protons while the latter involves both protons and neutrons.\footnote{The other difference is that, contrary to $\mathcal{E}^{(2)}_{2}$, $\mathcal{Q}^{(2)}_{2}$ involves the five components of the spherical tensor $Y^\mu_{2}(\Omega_{i})$, $\mu=-2,\ldots,+2$, such that it is a scalar with respect to rotations in three dimension. Interestingly, this would in fact only result in a prefactor difference for $J^\Pi=0^+$ states if the operators were summing over the same set of nucleons.} Thus, $\mathcal{Q}^{(2)}_{2}$ only probes proton correlations such that its expectation value, presently restricted to $J^\Pi=0^+$ ground-states,  can be expressed in terms of reduced electromagnetic quadrupole transition probabilities according to
\begin{equation}
    \langle \mathcal{Q}^{(2)}_{2} \rangle_{0^+_1} = \sum_{k} B(E2; 0^+_1 \rightarrow 2^+_k) \, . \label{decompokumarBEl}
\end{equation}
This testifies that this quadrupole collectivity measure can be reconstructed via low-energy experiments delivering quadrupole electromagnetic transitions to {\it all} $2^+$ excited states. In practice, only an approximate value associated with truncating the sum in Eq.~\eqref{decompokumarBEl} is effectively accessible for any given nucleus.

Within the classical RR model, the sum in Eq.~\eqref{decompokumarBEl} effectively reduces to a single transition within the ground-state band whose reduced probability is expressible in terms of the RR intrinsic quadrupole deformation $\beta_{20}$~\cite{RingSchuck,GreinerMaruhn}. As detailed in the SM, this eventually leads to
\begin{equation}
    \langle \mathcal{Q}^{(2)}_{2} \rangle^{\text{RR}}_{0^+_1} = \frac{9 Z^2e^2R_0^4}{16\pi^2} \beta_{20}^2 \; , \label{KumarRR}
\end{equation}
where $R_0\equiv 1.2 \, A^{1/3}$, i.e., the ground-state expectation value of Kumar operator $\mathcal{Q}^{(2)}_{2}$ is directly proportional to $\beta_{20}^2$, which is at variance with Eq.~\eqref{eq: mean squared anisotropy projected rigid rotor} for $\delta \epsilon^2_2$ (and thus for $\mathcal{E}^{(2)}_{2}$). Based on this model result, a quantity homogeneous to the square of a dimensionless quadrupole deformation parameter is introduced according to
\begin{equation}
    \label{eq: Beta parameter definition2}
    \mathcal{B}^2_{2}(\text{LE}) \equiv \frac{16\pi^2}{9 Z^2e^2R_0^4}  \langle \mathcal{Q}^{(2)}_{2} \rangle \; .
\end{equation}


\paragraph{Results.}

\begin{figure}[ht]
    \centering
    \includegraphics[width=\columnwidth]{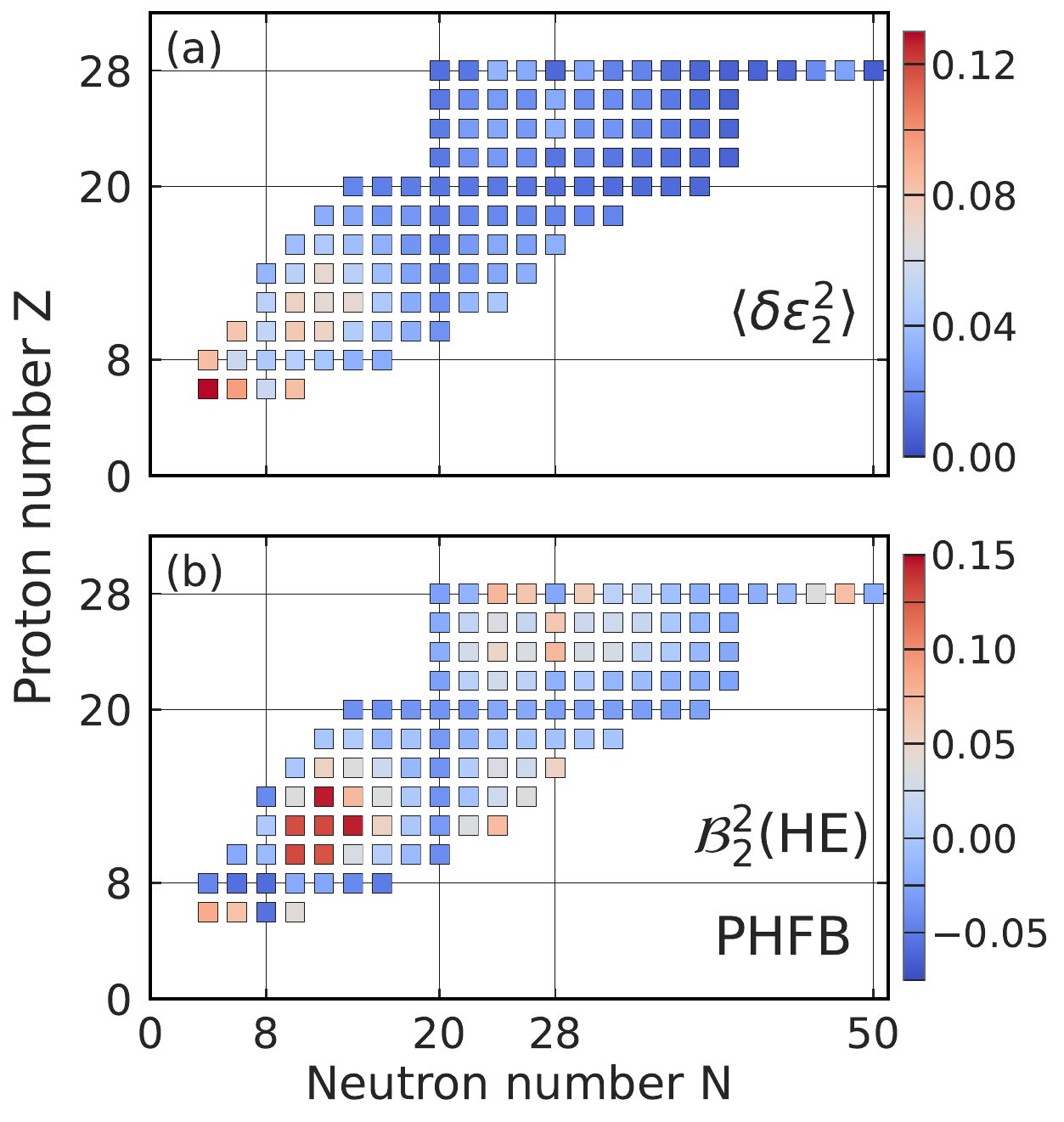}
    \caption{Projected HFB ground-state (a)  mean-square quadrupole anisotropy $\langle \delta \epsilon^2_2 \rangle$ and (b) its rescaled two-body contribution $\mathcal{B}^2_2(\text{HE})$ for even-even nuclei between carbon and nickel. Proton (neutron) magic numbers are indicated by horizontal (vertical) lines.}
    \label{fig:systematic}
\end{figure}

\begin{figure}[h!]
    \centering
    \includegraphics[width=\columnwidth]{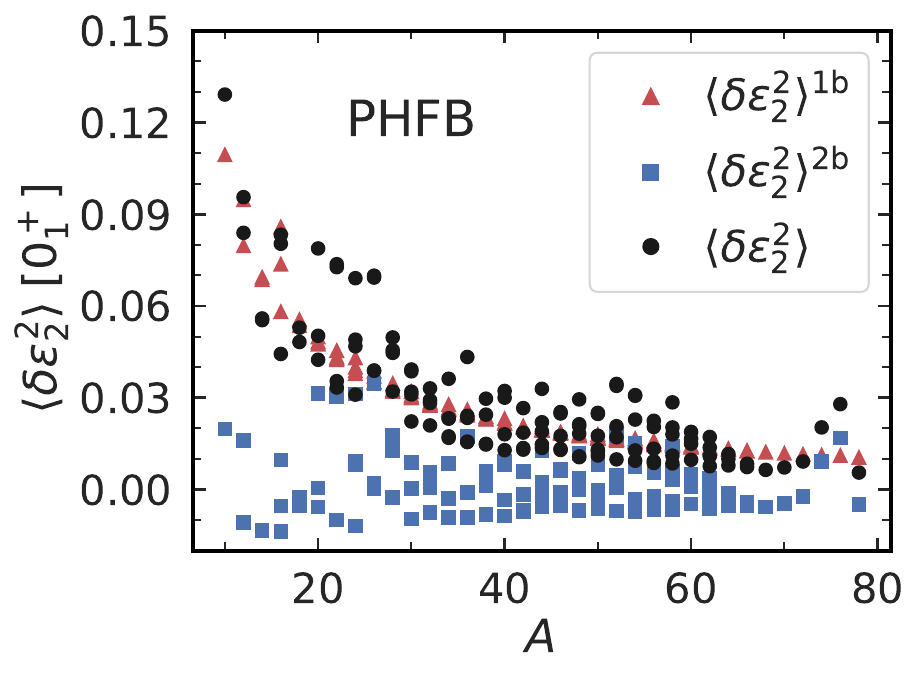}
    \caption{Projected HFB ground-state $\langle \delta \epsilon^2_2 \rangle$,  $\langle \delta \epsilon^2_2 \rangle^{1\text{b}}$ and  $\langle \delta \epsilon^2_2 \rangle^{2\text{b}}$ as a function of $A$ for even-even nuclei between carbon and nickel.}
    \label{fig: systematic comparison of 1- and 2-body parts of mean squared anisotropy}
\end{figure}

Panel (a) of Fig.~\ref{fig:systematic} displays the PHFB ground-state normalized mean-square quadrupole eccentricity $\langle \delta \epsilon^2_2 \rangle$ of even-even nuclei between carbon and nickel. This quantity is positive but particularly small for nuclei characterized by proton and/or neutron magic numbers. This total normalized mean-square quadrupole eccentricity displays mild fluctuations as a function of proton and neutron numbers and essentially decreases with A. The RR approximation to $\langle \delta \epsilon^2_2 \rangle$ in Eq.~\eqref{eq: mean squared anisotropy projected rigid rotor} suggests that this structureless decreasing behavior is driven by the trivial $1/A$ one-body contribution that dominates in the light nuclei under consideration. This is indeed numerically confirmed for our PHFB calculations in Fig.~\ref{fig: systematic comparison of 1- and 2-body parts of mean squared anisotropy} and in the SM. 

The key information about two-body quadrupole correlations lies in the two-body contribution $\langle \delta \epsilon^2_2\rangle^{2\text{b}}$, also shown in Fig.~\ref{fig: systematic comparison of 1- and 2-body parts of mean squared anisotropy}, and whose rescaled value $\mathcal{B}^2_2(\text{HE})$ [Eq.~\eqref{eq: Beta parameter definition1}] is displayed in panel (b) of Fig.~\ref{fig:systematic}. One observes more distinctive patterns such as small negative values along the $N$ or $Z=8,20$ semi-magic chains. Along the $N$ and $Z=28$ semi magic chains, $\mathcal{B}^2_2(\text{HE})$ is negative in doubly magic nuclei but becomes slightly positive in between them, testifying for increased quadrupole correlations. For nuclei with $8<N,Z<20$, especially around $^{20}$Ne ($Z=N=10$) large positive values of $\mathcal{B}^2_2(\text{HE})$ are predicted, signaling strong collective quadrupole correlations. While smaller in absolute than around $^{20}$Ne, positive values are predicted systematically between magic numbers.

\begin{figure}[t!]
\centering    \includegraphics[width=\columnwidth]{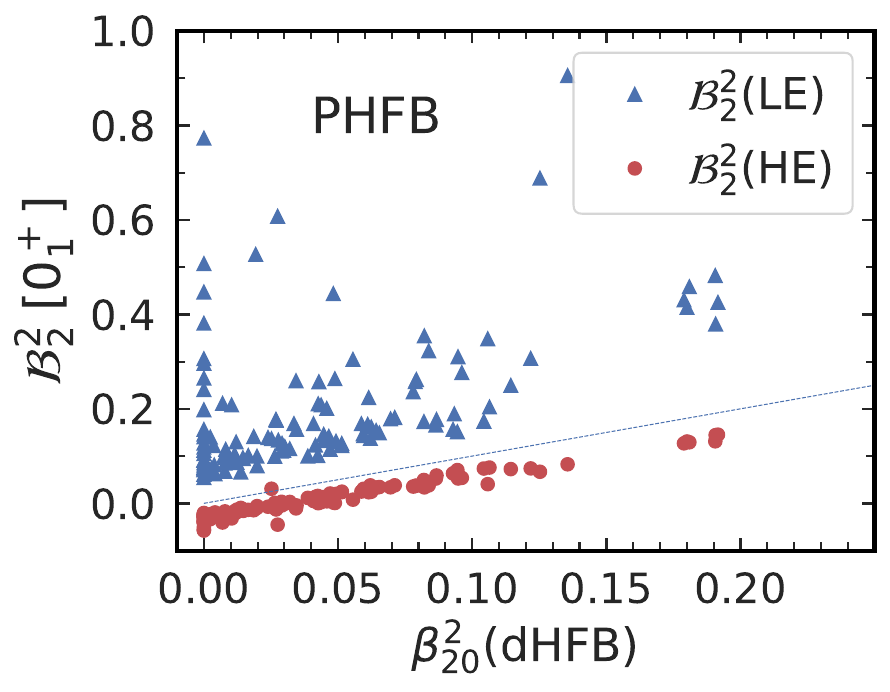}
    \caption{Rescaled PHFB two-body ground-state quadrupole mean-square eccentricity $\mathcal{B}^2_{2}(\text{HE})$ and Kumar quadrupole second moment $\mathcal{B}^2_{2}(\text{LE})$ against the squared intrinsic quadrupole deformation parameter $\beta^2_{20}$ of the underlying dHFB state. For consistency, the dHFB (PHFB) one-body radius is used to compute $\beta_{20}(\text{dHFB})$ ($\mathcal{B}^2_{2}(\text{LE})$) instead of $R_0$ using the connection detailed in the SM.}
    \label{fig: systematic comparison of Beta^2 versus standard beta^2}
\end{figure}

To elucidate these patterns and connect with the terminology of nuclear shapes and intrinsic deformation~\cite{Verney2025a}, PHFB $\mathcal{B}^2_2(\text{HE})$ values are displayed in Fig.~\ref{fig: systematic comparison of Beta^2 versus standard beta^2} against the squared intrinsic deformation of the underlying dHFB solution $| \Phi(\beta_{20},\beta_{30}) \rangle_{\text{dHFB}}$. Except for a negative offset, $\mathcal{B}^2_2(\text{HE})$ correlates essentially perfectly with $\beta_{20}^2(\text{dHFB})$. Using PHFB, i.e., a quantum rotor approximation, for $J^\Pi=0^+$ ground states, quadrupole two-body correlations are such that $\mathcal{B}^2_2(\text{HE})$ does characterize, up to an offset, the intrinsic quadrupole deformation. The offset from the first diagonal takes its origin in the negative value observed along semi-magic chains in Fig.~\ref{fig:systematic} and recovered in Fig.~\ref{fig: systematic comparison of Beta^2 versus standard beta^2} for ``spherical'' nuclei, i.e., nuclei with $\beta_{20}^2(\text{dHFB})=0$. This negative offset relates to quadrupolar correlations induced by the Pauli exclusion principle materialized by the exchange term in two-body density matrix~\cite{Blaizot2025a}, which is absent from the classical RR model employed to derive Eq.~\eqref{eq: mean squared anisotropy projected rigid rotor}. As shown in the SM, the magnitude of the offset decreases with $A$ and is essentially constant with $\beta_{20}^2(\text{dHFB})$, as can also be inferred from Fig.~\ref{fig: systematic comparison of Beta^2 versus standard beta^2}.

Figure~\ref{fig: systematic comparison of Beta^2 versus standard beta^2} also displays the low-energy-based parameter $\mathcal{B}^2_{2}(\text{LE})$. In spite of being extracted from the same PHFB quantum rotor approximation, $\mathcal{B}^2_{2}(\text{LE})$ does not correlate convincingly with the intrinsic deformation. This demonstrates that the traditional interpretation of the Kumar operator based on the RR model applied to $B(E2; 0^+_1 \rightarrow 2^+_k)$ is not operative and must be discarded.\footnote{Other limitations regarding the physical interpretation of Kumar invariants were pointed out in Ref.~\cite{Poves2020a}.} Not a priori separating the one and two-body parts of $\mathcal{Q}^{(2)}_{2}$, the RR analysis wrongly delivers $\langle \mathcal{Q}^{(2)}_{2} \rangle^{\text{RR}}_{0^+} \propto \beta_{20}^2$ [Eq.~\eqref{KumarRR}]. This leads to implicitly including the trivial $1/A$ one-body contribution into the definition of $\mathcal{B}^2_{2}(\text{LE})$, thus compromising the correlation with the actual intrinsic deformation in PHFB calculations in light nuclei. Contrarily, $\mathcal{B}^2_{2}(\text{HE})$ solely extracts two-body correlations and correlates perfectly with $\beta_{20}^2(\text{dHFB})$ in PHFB calculations. 

\begin{figure}[h!]
    \centering
    \includegraphics[width=\columnwidth]{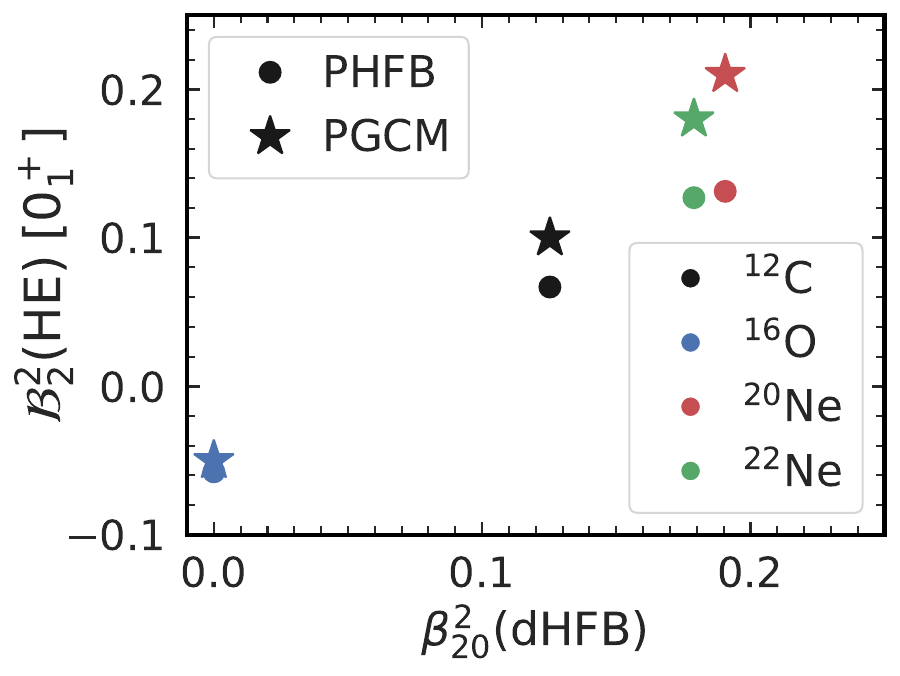}
    \caption{Same as Fig.~\ref{fig: systematic comparison of Beta^2 versus standard beta^2} for both PHFB and PGCM calculations of $^{12}$C, $^{16}$O and $^{20,22}$Ne.}
    \label{fig: PHFB-PGCM comparison of Beta^2 for selected nuclei}
\end{figure}

However, it must be clear that the PHFB approximation does not constitute a complete enough description of most nuclear ground states, i.e., PHFB is missing collective and non-collective correlations beyond those included in a quantum rotor approximation. To quantify the departure from the results obtained with PHFB, PHFB and PGCM $\mathcal{B}^2_{2}(\text{HE})$ values are plotted in Fig.~\ref{fig: PHFB-PGCM comparison of Beta^2 for selected nuclei} against $\beta^2_{20}(\text{dHFB})$ for the ground state of $^{12}$C, $^{16}$O and $^{20,22}$Ne. In addition to collective correlations associated with the rotation of a deformed body, present PGCM calculations include collective correlations induced by axial quadrupole and octupole shape fluctuations.

It is observed in Fig.~\ref{fig: PHFB-PGCM comparison of Beta^2 for selected nuclei} that such correlations increase with the size of the underlying dHFB intrinsic quadrupole deformation. Indeed, while the ``spherical'' $^{16}$O ($\beta^2_{20}(\text{dHFB})=0.00$) is very ``rigid'', the highly ``deformed'' $^{22}$Ne ($\beta^2_{20}(\text{dHFB})=0.18$) and $^{20}$Ne ($\beta^2_{20}(\text{dHFB})=0.19$) are much ``softer'', to the point that $\mathcal{B}^2_{2}(\text{HE})$ increases by about $38\%$ and $61\%$, respectively, compared to the PHFB value. Eventually, shape fluctuations do not induce a simple offset with respect to the rotor baseline. While numerically much more costly than PHFB calculations, it will be of interest to perform large-scale PGCM calculations to infer systematically the way shape fluctuations enlarge $\mathcal{B}^2_{2}(\text{HE})$ as a function of both $\beta^2_{20}(\text{dHFB})$ and $A$. 

Eventually, $\mathcal{B}^2_{2}(\text{HE})$ always captures both the negative offset due to Pauli's exclusion principle and the nucleus-dependent positive contribution originating from collective shape fluctuations in addition to the contribution matching the rotor intrinsic deformation, i.e., $\mathcal{B}^2_{2}(\text{HE})$ must not be blindly interpreted as the intrinsic deformation.\footnote{This statement is valid independently of the many-body method used to solve Schrödinger's equation.}. In fact, missing dynamical correlations must be added to the PGCM that can further move $\mathcal{B}^2_{2}(\text{HE})$ away from the quantum rotor intrinsic deformation value $\beta^2_{20}(\text{dHFB})$. While including such dynamical correlations to the PGCM is beyond the scope of the present letter, it can eventually be achieved perturbatively~\cite{Frosini2022a,Frosini2022b,Frosini2022c} 

\renewcommand{\arraystretch}{1.6} 
\begin{table}[t!]
\centering
\caption{Ground-state  rescaled quadrupole ($\mathcal{B}^2_{2}(\text{HE})$) and octupole ($\mathcal{B}^2_{2}(\text{HE})$) two-body mean-square eccentricities for $^{12}$C, $^{16}$O and $^{20,22}$Ne. (i)  Projected HFB, (ii) PGCM, (iii) NLEFT~\cite{Blaizot2025a} and (iv) QMC~\cite{Blaizot2025a}.}
\label{tab:example}
\setlength{\tabcolsep}{2pt} 
\begin{ruledtabular}
\begin{tabular}{c|cc|cc|cc|cc}
    & \multicolumn{2}{c}{$^{12}\mathrm{C}$}  & \multicolumn{2}{c}{$^{16}\mathrm{O}$}  & \multicolumn{2}{c}{$^{20}\mathrm{Ne}$} & \multicolumn{2}{c}{$^{22}\mathrm{Ne}$} \\  
        & $\mathcal{B}^2_2$  & $\mathcal{B}^2_3$ & $\mathcal{B}^2_2$  & $\mathcal{B}^2_3$ & $\mathcal{B}^2_2$  & $\mathcal{B}^2_3$ & $\mathcal{B}^2_2$  & $\mathcal{B}^2_3$ \\  \hline
i   & $+0.07$ & $+0.00$  & $-0.06$ & $+0.00$ &  $+0.13$ & $-0.07$ & $+0.13$ & $-0.07$ \\ \hline
ii  & $+0.10$ & $+0.01$  & $-0.05$ & $+0.18$ &  $+0.21$ & $+0.10$ & $+0.18$ & $-0.01$ \\ \hline
iii &         &          & $+0.04$ & $+0.12$ &  $+0.20$ & $+0.10$ &         &         \\ \hline
iv  &         &          & $-0.01$ & $+0.22$ &          &         &         &         \\ 
\end{tabular}
\end{ruledtabular}
\end{table}

Table~\ref{tab:example} lists $\mathcal{B}^2_{2,3}(\text{HE})$ values for the ground state of $^{12}$C, $^{16}$O and $^{20,22}$Ne. The large increase of $\mathcal{B}^2_{2}(\text{HE})$ in $^{20}$Ne from PHFB to PGCM due to shape fluctuations makes the PGCM result ($+0.21$) consistent with the  Nuclear Lattice Effective Field Theory (NLEFT) value ($+0.20$)~\cite{Blaizot2025a}. Such a situation extends to octupole correlations with fluctuations turning a moderately negative PHFB value of $\mathcal{B}^2_{3}(\text{HE})$ ($-0.07$) into a moderately  positive PGCM value ($+0.10$) in perfect agreement with NLEFT. These enhancement results from the intrinsic shape fluctuating into ``bowling-pin'' $\alpha$-clustered type configurations. Such an enhancement is reduced in $^{22}$Ne by the addition of two neutrons, leading to an essentially null PGCM $\mathcal{B}^2_{3}(\text{HE})$ value.

Similarly, $\mathcal{B}^2_{2,3}(\text{HE})$ PGCM values in $^{16}$O are in moderate agreement with NLEFT results but in good agreement with Quantum Monte Carlo (QMC) results~\cite{Blaizot2025a}. The discrepancy with NLEFT results may originate from the axial approximation used in the present work, which was not employed in the more complete calculations of Refs.~\cite{Giacalone2025a,Giacalone2025b}. Shape fluctuations significantly increase octupole correlations to reach $\mathcal{B}^2_{3}(\text{HE})=+0.18$, relatively close to the QMC value of $+0.22$. A similar analysis is provided in the SM  for $0^+_2$ excited states such as the Hoyle state in $^{12}$C.

Finally, one notices that $\mathcal{B}^2_{2}(\text{HE})$, which relates to the two-body part of $\langle \delta \epsilon^2_2 \rangle$ through Eq.~\eqref{eq: Beta parameter definition1}, is much larger in $^{20}$Ne than in $^{16}$O. Given that the one-body contribution to $\langle \delta \epsilon^2_2 \rangle$ is very similar in both nuclei, the excess of elliptic flow in $^{20}$Ne + $^{20}$Ne collisions compared to $^{16}$O + $^{16}$O, initially predicted in Refs.~\cite{Giacalone2025a,Giacalone2025b} and observed experimentally at the LHC \cite{CMS:2025tga,ALICE:2025luc,ATLAS:2025meas}, directly reflects the different nature of their two-body quadrupolar correlations. 

\paragraph{Conclusions and Outlook.} 

Characterizing  the correlated behavior of nucleons inside atomic nuclei constitutes a long-standing challenge both from an experimental point of view and from an interpretation one. 

The present work demonstrates via systematic {\it ab initio} nuclear structure calculations of light nuclei that two-particle correlations in the azimuthal distribution of final hadrons emitted in ultra-relativistic ultra-central ion-ion collisions can provide a meaningful imaging of nuclear ground-states. This imaging authorizes a robust interpretation in terms of  rotational and vibrational collective correlations. 
In light nuclei under study, the experimental observable however incorporates a trivial contribution overshadowing the two-body correlations of interest. 
It is thus necessary to combine results from different ion-ion collisions of well-chosen species to isolate, or at least magnify, genuine two-body correlations. This is what has been already achieved recently by taking the ratio of the $^{20}$Ne + $^{20}$Ne over $^{16}$O + $^{16}$O elliptic (triangular) flows to magnify quadrupolar (octupolar) correlations at play in both nuclei.

A future interesting development will consist of targeting specific three-particle correlations to isolate three-nucleon correlations in which additional nuclear structure information of interest leave their fingerprint.

\bigskip
\paragraph{Acknowledgements.} 
We thank J.-P.~Blaizot and G.~Giacalone for stimulating discussions as well as V. Som\`a for proofreading the manuscript. We also thank A.~Tichai for helping us benchmark our numerical implementation. The calculations were performed using computational resources from CCRT (TOPAZE and IRENE supercomputers). This work has received funding from the European Research Council under the European Union’s Horizon Europe Research and Innovation Programme (Grant Agreement No.\ 101162059).


\section{Supplemental Material}

\section{Eccentricity variance}

\subsection{First quantization formulation}

In the following, the position of a nucleon will be labeled by the spherical coordinates, $r$ and $\Omega$, as well as the cartesian coordinates, $x$ and $y$.

The one-body eccentricity operator of multipolarity $\ell>1$ is defined as
\begin{subequations}
\label{eq: SM 1-body eccentricity operator definition total}
\begin{align}
    \mathcal{E}^{(1)}_{\ell} &\equiv \sum_{i_1=1}^{A}  (x_{i_1} + iy_{i_1})^{\ell} \\
    &= \sum_{i_1=1}^{A} r_{i_1}^\ell e^{i \ell \phi_{i_1}} \sin^{\ell}\theta_{i_1} \\
    &= \sum_{i_1=1}^{A} \frac{r_{i_1}^\ell}{c_{\ell}} \, Y_\ell^{\ell}(\Omega_{i_1}) \; ,
\end{align}
\end{subequations}
where the normalization constant
\begin{equation}
    \label{eq: SM c_ell constant}
    c_{\pm \ell} = \frac{(\mp1)^{\ell}}{2^\ell l!}  \sqrt{\frac{(2\ell +1)!}{4\pi}} \; ,
\end{equation}
is such that $Y_\ell^{\pm \ell}(\Omega)$ is normalized on the unit sphere. Introducing for later use the one-body spherical tensor operators $E^{\pm \, (1)}_{\ell}\equiv \{E^{\pm \, (1)}_{\ell \mu}, \mu=-\ell,-\ell+1,...,\ell-1,\ell\}$ whose components are defined as
\begin{align}
\label{eq: SM 1-body E_lm operator 1st quant}
    E^{\pm \, (1)}_{\ell \mu} &= \sum_{i_1=1}^{A} \frac{r_{i_1}^\ell}{c_{\pm \ell}} \, Y_\ell^{\mu}(\Omega_{i_1}) \; ,
\end{align}
the eccentricity operators are nothing but 
\begin{equation}
    \mathcal{E}^{(1)}_{\pm \ell} \equiv E^{\pm \, (1)}_{\ell \pm \ell}\; .
\end{equation}

Given the fluctuation of the eccentricity 
\begin{equation}
\label{eq: SM 1-body eccentricity operator fluctuation}
    \delta \mathcal{E}^{(1)}_{\ell} \equiv \mathcal{E}^{(1)}_{\ell} - \langle \mathcal{E}^{(1)}_{\ell} \rangle \; ,
\end{equation} 
 where the expectation value is taken with respect to the many-body state of interest $| \Psi^{J\Pi}_k \rangle$, and the squared eccentricity operator
\begin{equation}
\label{eq: SM 2-body eccentricity operator definition}
\begin{aligned}
    \mathcal{E}^{(2)}_{\ell} &\equiv \mathcal{E}^{(1)}_{\ell} \mathcal{E}^{(1)}_{-\ell} ,
\end{aligned}
\end{equation}
the expectation value of the eccentricity variance reads as
\begin{equation}
    \label{eq: SM 2-body eccentricity operator expectation value}
    \langle \delta \mathcal{E}^{(2)}_{\ell} \rangle  \equiv \langle \delta \mathcal{E}^{(1)}_{\ell} \delta \mathcal{E}^{(1)}_{-\ell} \rangle  = \langle  \mathcal{E}^{(2)}_{\ell}\rangle - |\langle \mathcal{E}^{(1)}_{\ell}\rangle|^2 \; .
\end{equation}

\subsection{Decomposition of $\mathcal{E}^{(2)}_{\ell}$}

The operator $\mathcal{E}^{(2)}_{\ell}$ is the sum of a one-body operator and a two-body operator, i.e.,
\begin{align}
\label{eq: SM eccentricity operator definition 2nd quantization}
    \mathcal{E}^{(2)}_{\ell} \equiv \mathcal{E}^{(2)\, 1\text{b}}_{\ell} + \mathcal{E}^{(2)\, 2\text{b}}_{\ell} \; ,
\end{align} 
where 
\begin{subequations}
\begin{align}
    \mathcal{E}^{(2)\, 1\text{b}}_{\ell}  &\equiv \sum_{i_1= 1}^{A} \frac{r_{i_1}^{2\ell}}{c_{\ell}c_{-\ell}}\, Y_\ell^{\ell}(\Omega_{i_1}) \, Y_\ell^{-\ell}(\Omega_{i_1})  \\
    &= \sum_{i_1= 1}^{A} r_{i_1}^{2\ell} \sin^{2\ell}\theta_{i_1} \\
    &= \sum_{i_1= 1}^{A}  (x^2_{i_1} + y^2_{i_1})^\ell \, , \\
    \mathcal{E}^{(2)\, 2\text{b}}_{\ell} &\equiv \frac{1}{2} \sum_{i_1 \neq i_2 = 1}^{A} 2 \frac{r_{i_1}^\ell r_{i_2}^\ell}{c_{\ell}c_{-\ell}} \, Y_\ell^{\ell}(\Omega_{i_1}) \,   Y_\ell^{-\ell}(\Omega_{i_2}) \; ,
\end{align}
\end{subequations}

\subsection{Normalized eccentricity variance}

Eventually, one is interested in the normalized variance of the eccentricity~\cite{Duguet2025a}
\begin{equation}
\label{eq: SM mean squared anisotropy}
\begin{aligned}
    \langle \delta\epsilon^2_\ell\rangle &\equiv \frac{1}{2} \frac{\langle \delta \mathcal{E}^{(2)}_{\ell} \rangle}{\langle   R^{(1)}_{\ell}  \rangle^2} \\
    &\equiv  \frac{1}{2} \frac{\langle\mathcal{E}^{(2)}_{\ell}\rangle - |\langle \mathcal{E}^{(1)}_{\ell}\rangle|^2}{\langle   R^{(1)}_{\ell}   \rangle^2} \\
    &= \frac{1}{2} \frac{\langle \mathcal{E}^{(2)\, 1\text{b}}_{\ell} \rangle + \langle \mathcal{E}^{(2)\, 2\text{b}}_{\ell} \rangle - |\langle \mathcal{E}^{(1)}_{\ell}\rangle|^2}{ \langle R^{(1)}_{\ell} \rangle^2} \; ,
\end{aligned}
\end{equation}
where the one-body \emph{transverse radius} operator is defined as
\begin{equation}
\label{eq: SM transverse radius operator definition}
\begin{aligned}
    R^{(1)}_\ell &\equiv \sum_{i_1=1}^{A} (x^2_{i_1} + y^2_{i_1})^{\ell/2} = \sum_{i_1=1}^{A} r^\ell_{i_1} \sin^\ell \theta_{i_1}\; ,
\end{aligned}
\end{equation}
such that
\begin{equation}
\mathcal{E}^{(2)\, 1\text{b}}_{\ell} \equiv R^{(1)}_{2\ell} \; .
\end{equation}
For $J^{\pi} = 0^+$ states under present investigation, the mean eccentricity $\langle \mathcal{E}^{(1)}_{\ell}\rangle$ is zero such that the last term in the numerator of Eq.~\eqref{eq: SM mean squared anisotropy} is dropped, leading to the following simplified version
\begin{align}
    \label{eq: SM decompo eccentricity operator expectation value}
\langle \delta \epsilon^2_\ell\rangle_{0^+} &= \frac{1}{2} \frac{\langle \mathcal{E}^{(2)\, 1\text{b}}_{\ell} \rangle_{0^+}}{ \langle R^{(1)}_{\ell} \rangle^2_{0^+}} + \frac{1}{2} \frac{\langle \mathcal{E}^{(2)\, 2\text{b}}_{\ell} \rangle_{0^+}}{ \langle R^{(1)}_{\ell} \rangle^2_{0^+}} \nonumber \\
&\equiv \langle \delta \epsilon^2_\ell\rangle^{1\text{b}}_{0^+} + \langle \delta \epsilon^2_\ell\rangle^{2\text{b}}_{0^+} \; .
\end{align}

\subsection{Second quantization I}

Assuming a one-body basis $\mathcal{B}_1 = \{| i \rangle \}$, e.g., the spherical harmonic oscillator (sHO) basis,
\begin{equation}
    |i\rangle \equiv a^{\dagger}_{i} |0\rangle,
\end{equation}
where $|0\rangle$ denotes the physical vacuum and where the operator $a^{\dagger}_{i}$ creates a nucleon in state $| i \rangle$, the second-quantized form of the eccentricity operator reads as
\begin{align}
    \mathcal{E}^{(1)}_{\ell}  &= \sum_{ij} (\mathcal{\epsilon}^{(1)}_{\ell})_{ij}  \, a^{\dagger}_{i} a_{j}  \; ,  \label{2ndqecc}
\end{align}
with
\begin{equation}
    (\mathcal{\epsilon}^{(1)}_{\ell})_{ij} \equiv \langle i | \frac{r ^\ell}{c_{\ell}} \, Y_\ell^{\ell}(\Omega) | j \rangle  \, . \label{eccME}
\end{equation}

Given Eqs.~\eqref{2ndqecc}-\eqref{eccME}, the one-body and two-body components of $\mathcal{E}^{(2)}_{\ell}$ are obtained applying standard Wick's theorem~\cite{wick_evaluation_1950} with respect to $|0\rangle$ according with
\begin{subequations}
\begin{align}
    \mathcal{E}^{(2)\, 1\text{b}}_{\ell}  &\equiv \sum_{ij} (\Bar{\mathcal{\epsilon}}^{(2)}_{\ell})_{ij} \, a^{\dagger}_{i} a_{j}  \; ,\\
    \mathcal{E}^{(2)\, 2\text{b}}_{\ell} &\equiv \frac{1}{4} \sum_{ijkl} (\Bar{\mathcal{\epsilon}}^{(2)}_{\ell})_{ijkl} \, a^{\dagger}_{i} a^{\dagger}_{j} a_{l} a_{k} \; ,
\end{align}
\end{subequations}
with
\begin{subequations}
\label{eq: SM eccentricity matrix elements definition}
\begin{align}
     (\Bar{\mathcal{\epsilon}}^{(2)}_{\ell})_{ij} &= \sum_k (\mathcal{\epsilon}^{(1)}_{\ell})_{i k} (\mathcal{\epsilon}^{(1)}_{-\ell})_{k j} \, ,
\label{eq: 1-body eccentricity matrix elements definition} \\[1ex] 
(\Bar{\mathcal{\epsilon}}^{(2)}_{\ell})_{ijkl} &= 2[
(\mathcal{\epsilon}^{(1)}_{\ell})_{ik} (\mathcal{\epsilon}^{(1)}_{-\ell})_{jl} - (\mathcal{\epsilon}^{(1)}_{\ell})_{il} (\mathcal{\epsilon}^{(1)}_{-\ell})_{jk}] \; ,
\label{eq: 2-body eccentricity matrix elements definition} 
\end{align}
\end{subequations}
where the two-body matrix elements are antisymmetrized.

\subsection{Scalar component of $\mathcal{E}^{(2)}_{\ell}$}

Presently focusing on $J^{\pi}=0^+$ states, only the scalar component of the squared eccentricity operator is needed. To access it, one expands $\mathcal{E}^{(2)}_{\ell}$ into $L$-rank components according to
\begin{equation}
\label{eq: SM 2-body eccentricity operator decomposition to L}
    \mathcal{E}^{(2)}_{\ell} =  \sum_{L=0}^{2\ell} [\mathcal{E}^{(2)}_{\ell}]_{L0}
    \equiv \sum_{L=0}^{2\ell} C^{L0}_{\ell \ell \ell -\ell}[\Tilde{\mathcal{E}}^{(2)}_{\ell}]_{L0} \, ,
\end{equation}
with
\begin{equation}
\label{eq: SM 2-body eccentricity operator L-component}
\begin{aligned}
    [\Tilde{\mathcal{E}}^{(2)}_{\ell}]_{L0} &\equiv \sum_{i_1 i_2 = 1}^A \frac{r^{\ell}_{i_1}r^{\ell}_{i_2}}{c_\ell c_{-\ell}} [ Y^{\ell}_{\ell}(\Omega_{i_1}) Y^{-\ell}_{\ell}(\Omega_{i_2})]_{L0} \\
    &\equiv \sum_{i_1 i_2 = 1}^A \frac{r^{\ell}_{i_1}r^{\ell}_{i_2}}{c_\ell c_{-\ell}} \sum_{\mu=-\ell}^\ell C^{L0}_{\ell \mu \ell -\mu}  Y^{\mu}_{\ell}(\Omega_{i_1}) Y^{-\mu}_{\ell}(\Omega_{i_2}) \\
    &\equiv \sum_{\mu=-\ell}^\ell C^{L0}_{\ell \mu \ell -\mu}  E^{+ \, (1)}_{\ell \mu} E^{- \, (1)}_{\ell -\mu} \\
    &\equiv [E^{+ \, (1)}_{\ell} \times E^{- \, (1)}_{\ell}]_{L0} \; ,
\end{aligned}
\end{equation}
where $C^{LM}_{l_1 m_1 l_2 m_2}$ denotes Clebsch-Gordan (CG) coefficients.
Taking $l_1=l_2=m_1=-m_2$, the CG coefficient in Eq.~\eqref{eq: SM 2-body eccentricity operator decomposition to L} reads as 
\begin{equation}
\label{eq: SM CG coefficients maximal projection}
    C^{L0}_{\ell \ell \ell -\ell} = (2\ell)! \sqrt{\frac{2L + 1}{(2\ell + L + 1)!(2\ell - L)!}} \; ,
\end{equation}
The operator $[\Tilde{\mathcal{E}}^{(2)}_{\ell}]_{L0}$ is the sum of a one-body operator and of a two-body operator, i.e.,
\begin{align}
\label{eq: SM L-component operator in one and two}
    [\Tilde{\mathcal{E}}^{(2)}_{\ell}]_{L0} \equiv [\Tilde{\mathcal{E}}^{(2)}_{\ell}]^{1\text{b}}_{L0} + [\Tilde{\mathcal{E}}^{(2)}_{\ell}]^{2\text{b}}_{L0} \; ,
\end{align} 
with 
\begin{subequations}
\label{eq: SM rescaled L 1- and 2-body operators}
\begin{align}
[\Tilde{\mathcal{E}}^{(2)}_{\ell}]_{L0}^{1\mathrm{b}} &\equiv \sum_{i_1 = 1}^A   \frac{r^{2\ell}_{i_1}}{c_{\ell}c_{-\ell}} \sum_{\mu=-\ell}^\ell C^{L0}_{\ell \mu \ell -\mu}  Y^{\mu}_{\ell}(\Omega_{i_1}) Y^{-\mu}_{\ell}(\Omega_{i_1}) \nonumber \\
    &= \sum_{i_1 = 1}^A  \frac{r^{2\ell}_{i_1}}{c_{\ell}c_{-\ell}} \frac{2\ell + 1}{\sqrt{4\pi (2L+1)}} C^{L0}_{\ell 0 \ell 0} Y^{0}_{L}(\Omega_{i_1}) \; , \\
[\Tilde{\mathcal{E}}^{(2)}_{\ell}]^{2\text{b}}_{L0} &\equiv \frac{1}{2}\sum_{i_1 \neq i_2 = 1}^A 2\frac{r^{\ell}_{i_1}r^{\ell}_{i_2}}{c_\ell c_{-\ell}}  \nonumber \\ 
    &\quad \quad \times \sum_{\mu=-\ell}^\ell C^{L0}_{\ell \mu \ell -\mu}  Y^{\mu}_{\ell}(\Omega_{i_1}) Y^{-\mu}_{\ell}(\Omega_{i_2})\; ,
\end{align}
\end{subequations}
where the inverse Clebsch-Gordan series relation~\cite{Varshalovich} was employed for the one-body term.
For $L=0$, $Y_0^0(\Omega) = 1/4\pi$ and the needed CG coefficients read as
\begin{equation}
\label{eq: SM CG coefficients L=M=0}
    C^{00}_{l_1 m_1 l_2 m_2} = \delta_{l_1, l_2} \delta_{m_1,-m_2} \frac{(-1)^{l_1-m_1}}{\sqrt{2l_1+1}} \; ,
\end{equation}
such that the scalar one-body and two-body components read as
\begin{subequations}
\label{eq: SM rescaled L-0 1- and 2-body operators}
\begin{align}
    [\Tilde{\mathcal{E}}^{(2)}_{\ell}]_{00}^{\mathrm{1b}}
    &= \sum_{i_1 = 1}^A \frac{(-1)^{\ell}}{4\pi} \frac{\sqrt{2\ell + 1}}{c_{\ell} c_{-\ell}} r^{2\ell}_{i_1} \, , \\ 
    [\Tilde{\mathcal{E}}^{(2)}_{\ell}]_{00}^{\mathrm{2b}}&= \frac{1}{2}\sum_{i_1 \neq i_2 = 1}^A  \frac{2}{c_{\ell} c_{-\ell}} r^{\ell}_{i_1}r^{\ell}_{i_2} \nonumber \\
    &\quad\quad \times \sum_{\mu=-\ell}^\ell \frac{(-1)^{\ell - \mu}}{\sqrt{2\ell + 1}}  Y^{\mu}_{\ell}(\Omega_{i_1}) Y^{-\mu}_{\ell}(\Omega_{i_2})\, .
\end{align}
\end{subequations} 
Combining Eqs.~\eqref{eq: SM CG coefficients maximal projection} and~\eqref{eq: SM rescaled L-0 1- and 2-body operators} allows one to express the scalar ($L=0$) one-body and two-body components of the operator $\mathcal{E}^{(2)}_{\ell}$ as
\begin{subequations}
\begin{align}
    [\mathcal{E}^{(2)}_{\ell}]_{00}^{\mathrm{1b}}
    &= \sum_{i_1 = 1}^A \frac{1}{4\pi} \frac{(-1)^{\ell}}{c_{\ell} c_{-\ell}} r^{2\ell}_{i_1} \, , \\ 
    [\mathcal{E}^{(2)}_{\ell}]_{00}^{\mathrm{2b}}&= \frac{1}{2}\sum_{i_1 \neq i_2 = 1}^A \frac{2}{2\ell + 1} \frac{(-1)^{\ell}}{c_{\ell} c_{-\ell}} r^{\ell}_{i_1}r^{\ell}_{i_2} \nonumber \\
    &\quad\quad \times \sum_{\mu=-\ell}^\ell (-1)^{\mu}  Y^{\mu}_{\ell}(\Omega_{i_1}) Y^{-\mu}_{\ell}(\Omega_{i_2})\, ,
\end{align}
\end{subequations}
where the expression of $[\mathcal{E}^{(2)}_{\ell}]_{00}^{\mathrm{1b}}$ is the correct form of the erroneous Eq.~(38) in the SM of Ref.~\cite{Duguet2025a}. Specifying the above for multipolarities $\ell=2$ and $\ell=3$ of present interest provides
\begin{subequations}
\label{eq: SM 1- and 2-body eccentricity operator 0-component}
\begin{align}
    [\mathcal{E}^{(2)}_{2}]^{1\text{b}}_{00}
    &= \frac{8}{15} \sum_{i_1 = 1}^A  r^{4}_{i_1} \; , \\
    [\mathcal{E}^{(2)}_{2}]^{2\text{b}}_{00}
    &= \frac{32 \pi}{75} \sum_{i_1 \neq i_2 = 1}^A  r^{2}_{i_1} r^{2}_{i_2}
     \sum_{\mu=-2}^2 (-1)^{\mu} Y^{\mu}_{2}(\Omega_{i_1})  Y^{-\mu}_{2}(\Omega_{i_2})\; , \\
    [\mathcal{E}^{(2)}_{3}]^{1\text{b}}_{00}
    &= \frac{16}{35} \sum_{i_1 = 1}^A r^{6}_{i_1} \; , \\
    [\mathcal{E}^{(2)}_{3}]^{2\text{b}}_{00}
    &= \frac{64 \pi}{245} \sum_{i_1 \neq i_2 = 1}^A r^{3}_{i_1} r^{3}_{i_2}
     \sum_{\mu=-3}^3 (-1)^{\mu}  Y^{\mu}_{3}(\Omega_{i_1})  Y^{-\mu}_{3}(\Omega_{i_2})\; .
\end{align}
\end{subequations}

\subsection{Second quantization II}

Expressing the one-body operators $E^{\pm \, (1)}_{\ell \mu}$ in their second-quantized form
\begin{equation}
    \label{eq: SM E tensor components}
    \begin{aligned}
        E^{\pm \, (1)}_{\ell \mu} = \sum_{ij} (e^{\pm \, (1)}_{\ell \mu})_{ij}  \, a^{\dagger}_{i} a_{j}  \; , 
    \end{aligned}
\end{equation}
with 
\begin{equation}
    (e^{\pm \, (1)}_{\ell \mu})_{ij} \equiv \langle i | \frac{r^\ell}{c_{\pm \ell}} \, Y_\ell^{\mu}(\Omega) | j \rangle  \, , \label{eccME_v2}
\end{equation}
one observes that the one-body matrix elements of the eccentricity operator $(\epsilon^{(1)}_{\pm\ell})_{ij} \equiv (e^{\pm \, (1)}_{\ell \pm\ell})_{ij}$.
Given Eqs.~\eqref{eq: SM E tensor components} and~\eqref{eccME_v2} the one-body and two-body components of $[\Tilde{\mathcal{E}}^{(2)}_{\ell}]_{L0}$ are obtained via the application of standard Wick's theorem~\cite{wick_evaluation_1950} as
\begin{subequations}
\begin{align}
    [\Tilde{\mathcal{E}}^{(2)}_{\ell}]_{L0}^{1\mathrm{b}} &\equiv \sum_{ij} ([\Bar{\Tilde{\mathcal{\epsilon}}}^{(2)}_{\ell}]_{L0})_{ij} \, a^{\dagger}_{i} a_{j}  \; ,\\
    [\Tilde{\mathcal{E}}^{(2)}_{\ell}]_{L0}^{2\mathrm{b}} &\equiv \frac{1}{4} \sum_{ijkl} ([\Bar{\Tilde{\mathcal{\epsilon}}}^{(2)}_{\ell}]_{L0})_{ijkl} \, a^{\dagger}_{i} a^{\dagger}_{j} a_{l} a_{k} \; ,
\end{align}
\end{subequations}
with
\begin{subequations}
\label{eq:SM L-0 eccentricity me definition}
\begin{align}
     ([\Bar{\Tilde{\mathcal{\epsilon}}}^{(2)}_{\ell}]_{L0})_{ij} &= \sum_{\mu = -\ell}^{\ell}C^{L0}_{\ell \mu \ell -\mu}  \sum_k(e^{+ \, (1)}_{\ell \mu})_{i k} (e^{- \, (1)}_{\ell -\mu})_{k j} \, ,
\label{eq:SM L-0 eccentricity 1-b me definition} \\[1ex] 
    ([\Bar{\Tilde{\mathcal{\epsilon}}}^{(2)}_{\ell}]_{L0})_{ijkl} &= \sum_{\mu = -\ell}^{\ell} C^{L0}_{\ell \mu \ell -\mu} \nonumber \\
    & \times  \sum_k 2[
    (e^{+ \, (1)}_{\ell \mu})_{ik} (e^{- \, (1)}_{\ell -\mu})_{jl} - (e^{+ \, (1)}_{\ell \mu})_{il} (e^{- \, (1)}_{\ell -\mu})_{jk}] \; ,
\label{eq:SM L-0 eccentricity 2-b me definition} 
\end{align}
\end{subequations}
where the two-body matrix elements are antisymmetrized. Of course, the matrix elements of the one and two-body parts of $[\mathcal{E}^{(2)}_{\ell}]_{L0}$ are trivially obtained by multiplying the above with $C^{L0}_{\ell \ell \ell -\ell}$.


\subsection{PGCM mean-square eccentricity}

Given the form of the PGCM ansatz presently employed
\begin{equation}
| \Psi^{J\Pi}_k \rangle \equiv \sum_{\beta_{20}, \beta_{30}} f^{J\Pi}_{k}(\beta_{20},\beta_{30}) P^{J\Pi NZ} | \Phi(\beta_{20},\beta_{30}) \rangle \, , 
\label{SM PGCMansatz}
\end{equation}
the expectation of a given operator $O$ can be reduced to the computation of off-diagonal kernels of the form $\langle \Phi(p) | O | \Phi(q) \rangle$, where  $|\Phi(p)\rangle$ and $|\Phi(q)\rangle$ denote two different Bogoliubov states entering Eq.~\eqref{SM PGCMansatz}~\cite{Frosini2022a}.

Further introducing off-diagonal one-body density matrices~\cite{RingSchuck}
\begin{subequations}
\begin{align}
\label{eq: SM off-diagonal density matrices def}
    \rho_{s_1 r_1}(p,q) &\equiv \frac{\langle \Phi(p) | a^{\dagger}_{r_1} a_{s_1} | \Phi(q) \rangle}{\langle \Phi(p) | \Phi(q) \rangle} \;, \\
    \kappa_{s_1 s_2}(p,q) &\equiv \frac{\langle \Phi(p) | a_{s_2} a_{s_1} | \Phi(q) \rangle}{\langle \Phi(p) | \Phi(q) \rangle} \; , \\
    \Bar{\kappa}^*_{r_1 r_2}(p,q) &\equiv \frac{\langle \Phi(p) |  a^{\dagger}_{r_1} a^{\dagger}_{r_2} | \Phi(q) \rangle}{\langle \Phi(p) | \Phi(q) \rangle} \; . 
\end{align}
\end{subequations}
the expectation value of one-body operators such as $\mathcal{E}^{(2)\, 1\text{b}}_{\ell}$ or $R^{(1)}_\ell$ reads as, e.g.,
\begin{equation}
    \label{eq: SM 1-body mean-squared eccentricity kernels}
    \langle \Phi(p) | \mathcal{E}^{(2)\, 1\text{b}}_{\ell} | \Phi(q) \rangle = \sum_{ij} (\Bar{\mathcal{\epsilon}}^{(2)}_{\ell})_{ij} \, \rho_{ji}(p,q)\; .
\end{equation}
Employing off-diagonal Wick's theorem~\cite{balian_nonunitary_1969,porro_off-diagonal_2022}, the kernel of the two-body part of the mean-square eccentricity is computed according to
\begin{widetext}
\begin{equation}
\begin{split}
\label{eq: SM 2-body mean-squared eccentricity kernels}
    \langle \Phi(p) | \mathcal{E}^{(2)\, 2\text{b}}_{\ell} | \Phi(q) \rangle &= \frac{1}{4} \sum_{ijkl} (\Bar{\mathcal{\epsilon}}^{(2)}_{\ell})_{ijkl} [\rho_{ki}(p,q)\rho_{lj}(p,q) - \rho_{kj}(p,q)\rho_{li}(p,q) + \Bar{\kappa}^*_{ij}(p,q) \kappa_{kl}(p,q) ] \\
    &=  \mathrm{Tr}[\mathcal{\epsilon}^{(1)}_{\ell} \rho(p,q)] \mathrm{Tr}[\mathcal{\epsilon}^{(1)}_{-\ell} \rho(p,q)] - \mathrm{Tr}[\mathcal{\epsilon}^{(1)}_{\ell} \rho(p,q) \mathcal{\epsilon}^{(1)}_{-\ell} \rho(p,q)] \\
    &\quad + \frac{1}{2} \mathrm{Tr}[\mathcal{\epsilon}^{(1)}_{\ell} \kappa(p,q) (\mathcal{\epsilon}^{(1)}_{-\ell})^{T} \Bar{\kappa}^{\dagger}(p,q)] - \frac{1}{2} \mathrm{Tr}[\mathcal{\epsilon}^{(1)}_{\ell} \kappa^{T}(p,q) (\mathcal{\epsilon}^{(1)}_{-\ell})^{T} \Bar{\kappa}^{\dagger}(p,q)] \\
    &= \mathrm{Tr}[\mathcal{\epsilon}^{(1)}_{\ell} \rho(p,q)] \mathrm{Tr}[\mathcal{\epsilon}^{(1)}_{-\ell} \rho(p,q)] - \mathrm{Tr}[\mathcal{\epsilon}^{(1)}_{\ell} \rho(p,q) \mathcal{\epsilon}^{(1)}_{-\ell} \rho(p,q)]
    + \mathrm{Tr}[\mathcal{\epsilon}^{(1)}_{\ell} \kappa(p,q) (\mathcal{\epsilon}^{(1)}_{-\ell})^{T} \Bar{\kappa}^{\dagger}(p,q)]
\end{split}
\end{equation}
\end{widetext}
where advantage was taken of the separable character of the two-body matrix elements (Eq.~\eqref{eq: 2-body eccentricity matrix elements definition}) as well as of the antisymmetry of the density $\kappa$ to obtain an expression that is much less costly to compute than for a general two-body operator.


\subsection{Classical rigid rotor interpretation}

The interpretation of the normalized mean-squared eccentricity $\langle \delta \epsilon^2_\ell\rangle_{0^+}$ within the classical rigid-rotor (RR) model allows for insightful conclusions regarding the nature of multi-particle correlations~\cite{Duguet2025a}. 

In its present version, the starting point is an axially-symmetric intrinsic local one-body density given by a (normalized) Gaussian profile of the form
\begin{equation}
    \label{eq: SM diagonal intrinsic one-body density gaussian form1}
    \begin{aligned}
    \rho^{(1)}_{\mathrm{intr}}(\mathbf{r}) 
    &\equiv A \frac{e^{-r^2/2R(\theta)^2}}{(2\pi)^{3/2}R^{3}}  \; ,
    \end{aligned}
\end{equation}
where the nuclear surface is expanded in terms of axial quadrupole and octupole deformations according to
\begin{equation}
    \label{eq: SM axial nuclear surface expansion}
    R(\theta) \equiv R[1 + \beta_{20}Y_2^0(\theta) + \beta_{30}Y_3^0(\theta)] ,
\end{equation}
and where $\theta$ is the angle in the intrinsic frame with respect to the symmetry axis.

Expanding the angular dependence in the intrinsic one-body density according to
\begin{equation}
    \label{eq: SM diagonal intrinsic one-body density gaussian form2}
    \begin{aligned}
    \rho^{(1)}_{\mathrm{intr}}(\mathbf{r}) 
     &\approx A \frac{e^{-r^2/2R^2}}{(2\pi)^{3/2}R^{3}}  [1 + \frac{r^2}{R^2} \beta_{20}Y_2^0(\theta) + \frac{r^2}{R^2}\beta_{30}Y_3^0(\theta)] \; ,
    \end{aligned}
\end{equation}
the local one-body density of the $0^+_1$ ground state reads in the laboratory frame as
\begin{equation}
    \label{eq: SM diagonal lab-frame one-body density}
    \begin{aligned}
    \rho^{(1)}_{\mathrm{lab}}(\mathbf{r}) 
    \equiv \frac{1}{4\pi} \int_{\Omega_{\mathrm{or}}}  
       \rho^{(1)}_{\mathrm{intr}}(R_{\Omega_{\rm or}}^{-1} \mathbf{r})  =  A \frac{e^{-r^2/2R^2}}{(2\pi)^{3/2}R^{3}}  \; ,
    \end{aligned}
\end{equation}
where $\Omega_{\mathrm{or}}$ denotes Euler angles $(\alpha, \beta)$ stipulating the orientation of the intrinsic deformation in space and $R_{\Omega_{\rm or}}^{-1}$ represents a rotation of the coordinate system from the laboratory frame to the intrinsic frame. The pre-factor $A$ ensures that  the integration of this model local one-body density over space is equal to $A$. 

Calculating the expectation value of the two-body part of the squared eccentricity operator requires the local two-body density, which is obtained in the classical RR model according to\footnote{Because of its classical character, the RR model only consider the direct part of the two-body density, which is thus not antisymmetric under the exchange of the two particles coordinates. As a result, its integration over space is equal to $A^2$ instead of $A(A-1)$ as it would be the case for a quantal two-body density. }\footnote{The fact that $\rho^{(2)}_{\mathrm{lab}}(\mathbf{r}_1, \mathbf{r}_2) $ has a contribution of the type $\beta_{\ell 0}^2 r_1^\ell r_2^\ell \cos(\ell(\phi_1-\phi_2))$ was argued for the first time in Ref.~\cite{Blaizot:2025scr} for a number of test cases.}
\begin{align}
    \label{eq: SM diagonal lab-frame two-body density}
    \rho^{(2)}_{\mathrm{lab}}(\mathbf{r}_1, \mathbf{r}_2) 
    &\equiv \frac{1}{4\pi} \int_{\Omega_{\mathrm{or}}}  
       \rho^{(1)}_{\mathrm{intr}}(R_{\Omega_{\rm or}}^{-1} \mathbf{r}_1) \rho^{(1)}_{\mathrm{intr}}(R_{\Omega_{\rm or}}^{-1} \mathbf{r}_2)  \nonumber \\
    &= A^2 \frac{e^{-(r^2_1 + r^2_2)/2R^2}}{32\pi^4 R^6}  \times [4\pi \\ 
    &\quad + \frac{r^2_1 r^2_2}{R^4}\beta^2_{20} P_{2}(\cos\zeta) + \frac{r^2_1 r^2_2}{R^4}\beta^2_{30} P_{3}(\cos\zeta)], \nonumber
\end{align}
where $P_\ell$ denotes Legendre polynomials and where
\begin{equation}
    \cos\zeta \equiv \frac{\mathbf{r}_1 \cdot \mathbf{r}_2}{r_1 r_2}\; .
\end{equation}

Evaluating Eq.~\eqref{eq: SM mean squared anisotropy} using Eqs.~\eqref{eq: SM diagonal lab-frame one-body density} and \eqref{eq: SM diagonal lab-frame two-body density} leads to
\begin{equation}
\label{eq: SM mean squared anisotropy rigid rotor}
\begin{aligned}
    \langle \delta \epsilon^2_2 \rangle^{\text{RR}}_{0^+_1} = \frac{1}{A} + \frac{3}{4\pi} \beta^2_{20}\; ,\quad
    \langle \delta \epsilon^2_3 \rangle^{\text{RR}}_{0^+_1} = \frac{16}{3\pi A} + \frac{2048}{245\pi^3} \beta^2_{30}\; .
\end{aligned}
\end{equation}
Based on this model result, a quantity homogeneous to the square of a dimensionless deformation parameter is introduced according to~\cite{Blaizot2025a}
\begin{equation}
    \label{eq: SM Beta parameter definition1}
    \mathcal{B}^2_{\ell}(\text{HE}) \equiv \frac{4\pi}{3} \langle \delta \epsilon^2_\ell\rangle^{2\text{b}} \; .
\end{equation}

\section{Kumar operator}

\subsection{Definition}

Kumar's operator of order $n$ and multipolarity $\ell$ is defined as~\cite{Kumar1972a}
\begin{equation}
    \label{eq: SM Kumar n-body operator definition}
    \mathcal{Q}^{(n)}_{\ell} \equiv ([Q^{(1)}_{\ell} \times Q^{(1)}_{\ell} ... \times Q^{(1)}_{\ell}]_{\ell} \cdot Q^{(1)}_{\ell})\; ,
\end{equation}
where  the one-body electric operator $Q^{(1)}_{\ell}$ of multipolarity $\ell$ is an irreducible tensor of rank $\ell$ whose $Q^{(1)}_{\ell \mu}$ components are given by
\begin{equation}
    \label{eq: SM quadupole operator definition}
    Q^{(1)}_{\ell \mu} \equiv \sum_{i_1=1}^{A} e_i  r^\ell_{i_1} Y^{\mu}_{\ell}(\Omega_{i_1})\;.
\end{equation}
The operator $\mathcal{Q}^{(n)}_{\ell}$ is a scalar operator where the term in square brackets with subscript $\ell$ signifies a tensor product series of rank $\ell$, and the entire expression within parentheses denotes a scalar product. 
In particular, the second order Kumar operator is expressed as a simple tensor scalar product
\begin{align}
    \label{eq: SM Kumar 2-body operator definition}
    \mathcal{Q}^{(2)}_{\ell} &\equiv (Q^{(1)}_{\ell} \cdot Q^{(1)}_{\ell}) \nonumber \\
    &\equiv \sum_{\mu=-\ell}^{\ell} (-1)^{\mu} Q^{(1)}_{\ell \mu} Q^{(1)}_{\ell -\mu} \\
    &\equiv \sum_{i_1 i_2 = 1}^A  e_{i_1} e_{i_2} r^{\ell}_{i_1} r^{\ell}_{i_2}  \sum_{\mu=-\ell}^{\ell} (-1)^{\mu}  Y^{\mu}_{\ell}(\Omega_{i_1})  Y^{-\mu}_{\ell}(\Omega_{i_2})\; . \nonumber
\end{align}
Clearly, the sole difference (up to a prefactor) between $\mathcal{Q}^{(2)}_{\ell}$ and $[\mathcal{E}^{(2)}_{\ell}]_{00}$ relates to the fact that the former only involve protons whereas the latter involves both protons and neutrons.

Probing only proton correlations, $\mathcal{Q}^{(2)}_{\ell}$ can be rewritten in terms of reduced electromagnetic transition probabilities, e.g., the expectation value of the quadrupole operator $\mathcal{Q}^{(2)}_{2}$ in the $J^{\pi} = 0^+$ ground state reads as
\begin{equation}
\label{eq: SM Kumar expectation value wrt BE2}
    \langle \mathcal{Q}^{(2)}_{2} \rangle_{0^+_1} = \sum_{k} B(E2; 0^+_1 \rightarrow 2^+_k) \; ,
\end{equation}
which makes clear that this correlations measure can be accessed experimentally by summing quadrupole electromagnetic transition probabilities to all $2^+$ excited states. In practice, however, only an approximate value is obtained through the truncation of the sum in Eq.~\eqref{eq: SM Kumar expectation value wrt BE2}.

\subsection{Classical rigid rotor interpretation}

It is customary, from a theoretical perspective, to interpret the ground-state expectation value of Kumar operators within the framework of simple collective models, such as the classical RR model. The assumption of the nucleus being a pure axial RR results in the reduction of the sum in Eq.~\eqref{eq: SM Kumar expectation value wrt BE2} into a single transition within the ground-state band. Combining this with the fact that the RR model allows one to express $ B(E2; 0^+_1 \rightarrow 2^+_1)$  in terms of the intrinsic quadrupole moment $Q_0$ of the band, eventually leads to
\begin{equation}
\begin{aligned}
    \langle \mathcal{Q}^{(2)}_{2} \rangle^{\text{RR}}_{0^+_1} &= B(E2; 0^+_1 \rightarrow 2^+_1) \\
    &= \frac{5}{16\pi}Q^2_0 \\
    &= \frac{9}{16\pi^2}Z^2 e^2 R^4_0 \beta^2_{20}\; ,
\end{aligned}
\end{equation}
where the definition $Q_0 \equiv \sqrt{\frac{16\pi}{5}} \frac{3}{4\pi}Z e R^2_0\beta_{20}$ was used.

Thus, within the classical RR model, the ground-state expectation value of $\mathcal{Q}^{(2)}_{2}$ is directly proportional to the square of the intrinsic deformation parameter $\beta_{20}$. Based on this model result, a quantity homogeneous to the square of a dimensionless quadrupole deformation parameter is thus introduced according to
\begin{align}
    \label{eq: SM Beta parameter definition2}
    \mathcal{B}^2_{2}(\text{LE}) &\equiv \frac{16\pi^2}{9 Z^2e^2R_0^4}  \langle \mathcal{Q}^{(2)}_{2} \rangle \nonumber \\
    &= \frac{16\pi^2A^2}{25Z^2e^2} \frac{\langle \mathcal{Q}^{(2)}_{2} \rangle}{\langle r^2 \rangle^2} \; ,
\end{align}
where the empirical radius $R^2_0 = 1.2A^{1/3}$ has been replaced by the computed PHFB or PGCM  (non-normalized\footnote{Mass mean-squared radii compared to experimental data include a normalization; e.g., the proton number for the point proton mean-square radius or the mass number for the mass mean-square radius. No such normalization is used in the present study when defining the expectation value of a squared radius operator.}) mean-square radius $\langle r^2 \rangle$. Doing so, the connection between the radius $R_0$ of a uniform sphere and the mean-square radius is given by
\begin{equation}
    \langle r^2 \rangle = \frac{3}{5} A R^2_0 \; .
\end{equation}


\section{Numerical results}

The present section extends the analysis of the numerical results presented in the bulk of the paper regarding the systematic study of the normalized mean-square eccentricity at the PHFB and PGCM levels.

\subsection{Technical details}

The systematic analysis of even-even nuclei between carbon and nickel is performed within the PHFB approximation based on the deformed HFB minimum and the EM1.8/2.0 chiral Hamiltonian~\cite{Hebeler2011a}. The three-body interaction is approximated via the rank-reduction method developed in Ref.~\cite{frosini2021medium}. A one-body sHO basis is employed, characterized by the frequency $\hbar \omega = 12$ MeV, known to be optimal for nuclei in the light to mid-mass regime. All single-particle basis states up to $e_{\mathrm{max}} = 10$ are included, while the representation of the initial three-body interaction operator is further restricted to three-body states up to $e_{\mathrm{3max}} = 24$.

\begin{figure}
    \centering
    \includegraphics[width=\columnwidth]{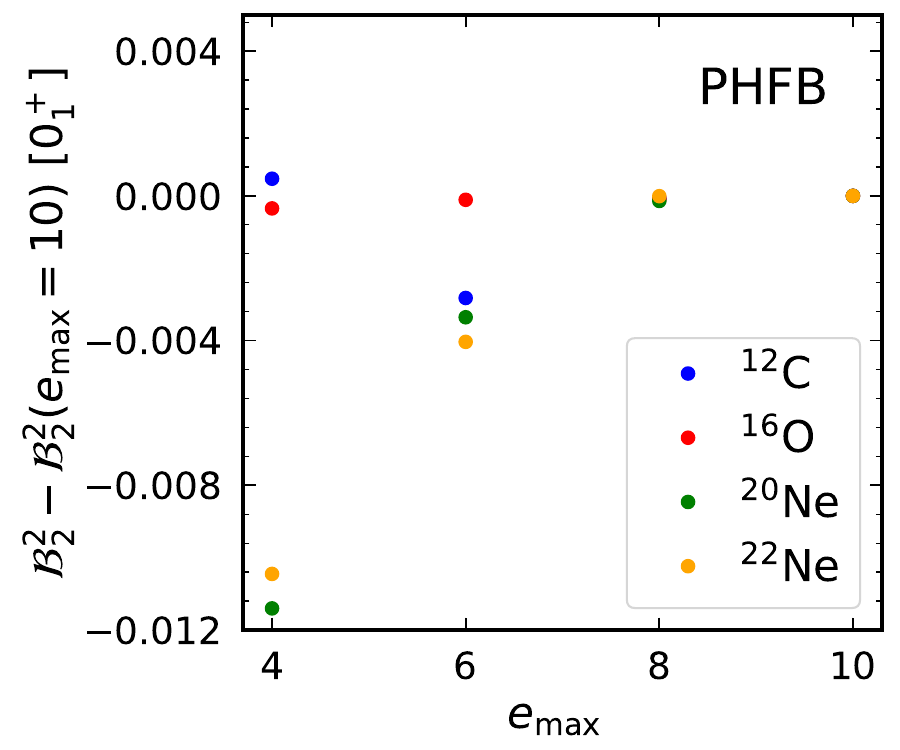}
    \caption{Convergence plot of $\mathcal{B}^2_2(\mathrm{HE})$ in $^{12}$C, $^{16}$O and $^{20,22}$Ne as a function of the one-body basis dimension. Calculations are performed within the PHFB approximation.}
    \label{fig: SM B2_sq cvg PHFB}
\end{figure}

\begin{figure}
    \centering
    \includegraphics[width=\columnwidth]{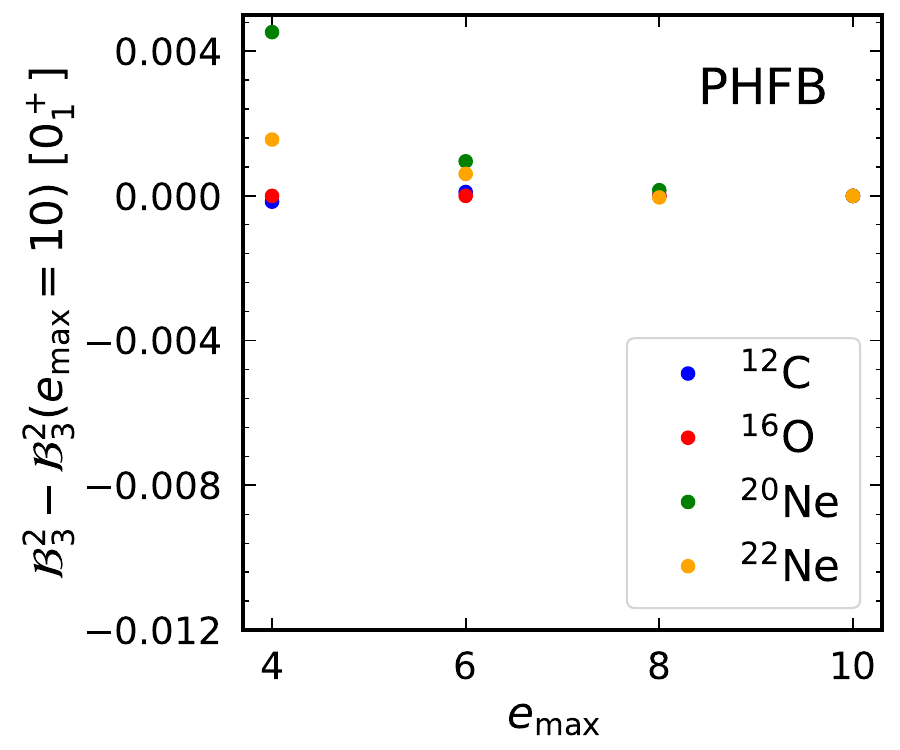}
    \caption{Same as Fig.~\ref{fig: SM B2_sq cvg PHFB} for $\mathcal{B}^2_3(\mathrm{HE})$.}
    \label{fig: SM B3_sq cvg PHFB}
\end{figure}

The nuclei $^{12}$C, $^{16}$O and $^{20,22}$Ne are further studied via PGCM calculations based on axially deformed Bogoliubov states~\cite{Frosini2022a,Frosini2022b,Frosini2022c}. To achieve an improved treatment of pairing correlations, the reference states entering the PGCM ansatz are generated using variation after particle-number projection, with subsequent projection  onto good angular momentum and parity. The single-particle basis is further restricted to $e_{\mathrm{max}} = 8$. As demonstrated  in Figs.~\ref{fig: SM B2_sq cvg PHFB} and~\ref{fig: SM B3_sq cvg PHFB} for PHFB calculations, the rescaled two-body mean-square eccentricities  $\mathcal{B}^2_{2,3}(\mathrm{HE})$ of present interest are well converged at $e_{\mathrm{max}} = 8$.

\subsection{One-body contribution $\langle \delta \epsilon^2_2\rangle^{\mathrm{1b}}$}

\begin{figure}[ht]
    \centering
    \includegraphics[width=\columnwidth]{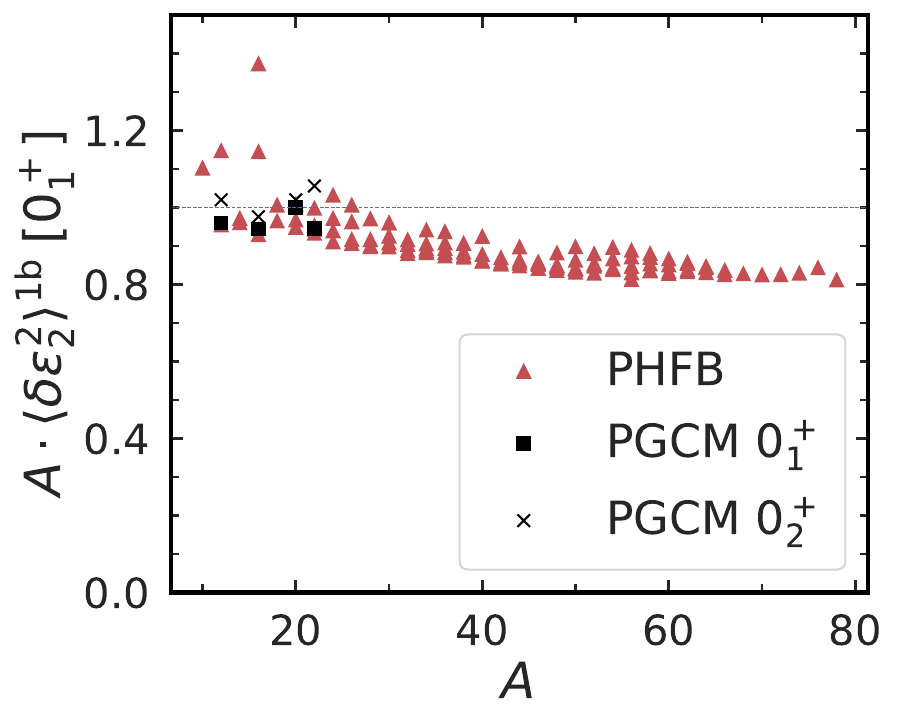}
    \caption{Ground-state product $\langle \delta \epsilon^2_2\rangle^{\mathrm{1b}}A$ from PHFB and PGCM calculations as a function of the mass number. }
    \label{fig: SM A times one-body anisotropy wrt A}
\end{figure}

As stipulated by Eq.~\eqref{eq: SM mean squared anisotropy rigid rotor}, the classical RR model suggests that the one-body contribution to the normalized mean-squared eccentricity $\langle \delta \epsilon^2_2\rangle^{\mathrm{1b}}$ is independent of the intrinsic deformation and equal to $1/A$. Figure~\ref{fig: SM A times one-body anisotropy wrt A} does display the product $A\langle \delta \epsilon^2_2\rangle^{\mathrm{1b}}$ from systematic PHFB calculations of nuclear ground states. The product varies only very moderately with $A$, following quite closely the RR approximation. Clearly, the few PGCM points visible on the figure remain consistent with this observation.

\begin{figure}[ht!]
    \centering
    \includegraphics[width=\columnwidth]{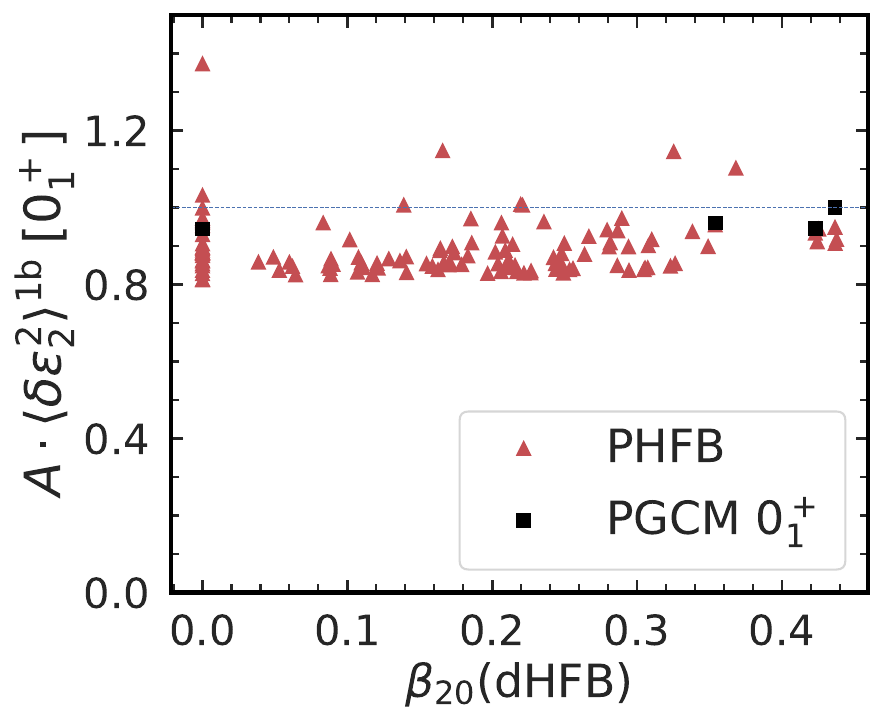}
    \caption{Same as Fig.~\ref{fig: SM A times one-body anisotropy wrt A} as a function of the intrinsic axial quadrupole deformation $\beta_{20}$ of the underlying dHFB state. PGCM results are plotted using the same $\beta_{20}$ values as the PHFB calculations.}
    \label{fig: SM A times one-body anisotropy wrt beta}
\end{figure}

In Fig.~\ref{fig: SM A times one-body anisotropy wrt beta}, the independence of the product $A\langle \delta \epsilon^2_2\rangle^{\mathrm{1b}}$ with respect to the intrinsic deformation is further confirmed. One observes that the shape fluctuations included in the PGCM wave-function ansatz do not modify this property.

\subsection{Two-body contribution $\langle \delta \epsilon^2_2\rangle^{\mathrm{2b}}$}

\begin{figure}[ht!]
    \centering
    \includegraphics[width=\columnwidth]{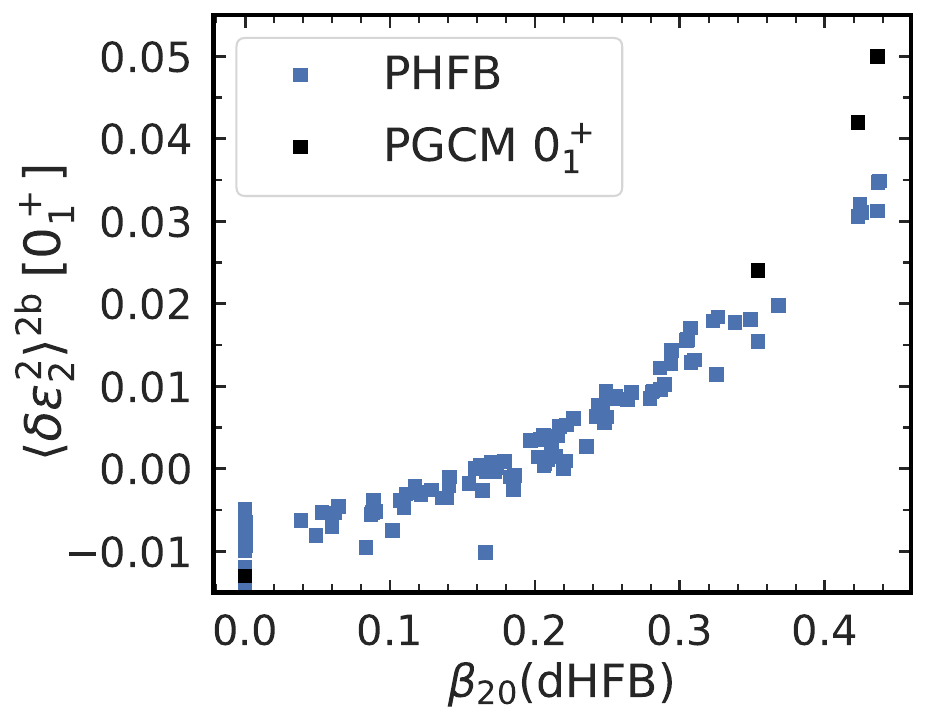}
    \caption{Ground-state value of $\langle \delta \epsilon^2_2\rangle^{\mathrm{2b}}$  from  PHFB and PGCM calculations as a function of the intrinsic axial quadrupole deformation $\beta_{20}$ of the underlying dHFB state. PGCM results are plotted using the same $\beta_{20}$ values as the PHFB calculations.}
    \label{fig: SM two-body anisotropy wrt beta}
\end{figure}

As discussed in the body of the paper, the RR model predicts the two-body contribution $\langle \delta \epsilon^2_2\rangle^{\mathrm{2b}}$ to the normalized mean-square eccentricity to scale with the square of the deformation parameter $\beta_{20}$ (Eq.~\eqref{eq: SM mean squared anisotropy rigid rotor}). This is indeed what is obtained for a quantum rotor, i.e., within the PHFB approximation, as shown in Fig.~\ref{fig: SM two-body anisotropy wrt beta}. In addition, a negative offset, not predicted by the classical rotor, is seen to shift that parabola. This offset takes its origin in the Pauli exclusion principle materialized by the exchange term in the two-body density matrix~\cite{Blaizot2025a} and is indeed absent from the classical RR model used to derive Eq.~\eqref{eq: SM mean squared anisotropy rigid rotor}. Because the (quantum) rotor character of the many-body state is altered by shape fluctuations, the PGCM results depart slightly from the PHFB trend.

\begin{figure}[ht!]
    \centering
    \includegraphics[width=\columnwidth]{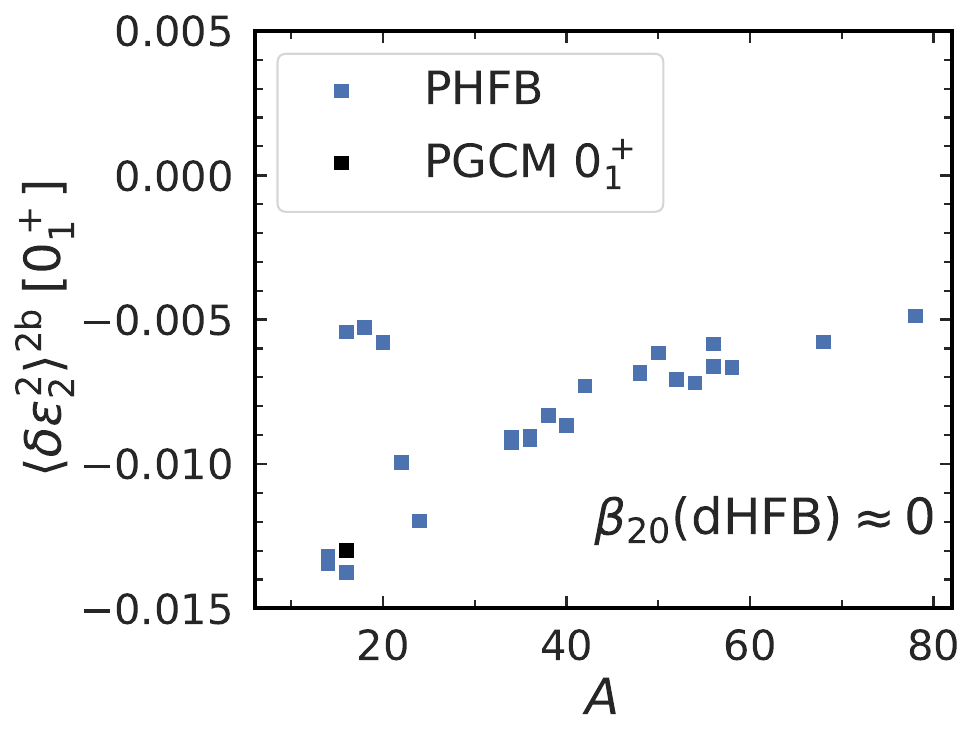}
    \caption{Ground-state value of $\langle \delta \epsilon^2_2\rangle^{\mathrm{2b}}$  from PHFB and PGCM calculations as a function of $A$ for nearly ``spherical'' nuclei characterized by $\beta_{20}^2(\text{dHFB}) \approx 0$.}
    \label{fig: SM two-body anisotropy offset at beta=0}
\end{figure}

Isolating $\langle \delta \epsilon^2_2\rangle^{\mathrm{2b}}$ for ``spherical'' nuclei characterized by $\beta_{20}^2(\text{dHFB}) \approx 0$, Fig.~\ref{fig: SM two-body anisotropy offset at beta=0} shows that the magnitude of the negative offset decreases with $A$. This particular property needs to be confirmed analytically~\cite{blaizot26}.

\begin{figure}
    \centering
    \includegraphics[width=\columnwidth]{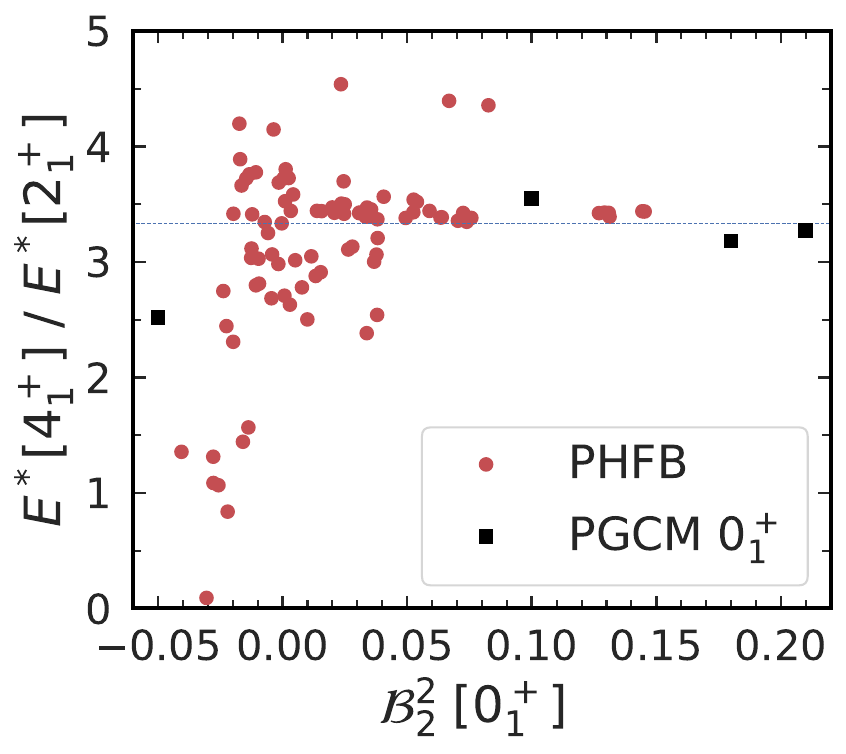}
    \caption{Ratio $R_{4/2} \equiv E^*(4^+_1)/E^*(2^+_1)$ of PHFB and PGCM $4^+_1$ to $2^+_1$ excitation energies as a function of the two-body normalized component $\mathcal{B}^2_2(\text{HE})$. The horizontal blue dotted line denotes the rigid rotor value, equal to 10/3.}
    \label{fig: SM R_42 ratio wrt B2_sq}
\end{figure}

A hallmark of a rigid rotor is the characteristic energy ratio $R_{4/2} \equiv E^*(4^+_1)/E^*(2^+_1) =10/3$, with $E^*$ denoting the excitation energy. Figure~\ref{fig: SM R_42 ratio wrt B2_sq} displays the $R_{4/2}$ ratio from our PHFB and PGCM calculations as a function of $\mathcal{B}^2_2(\mathrm{HE})$. For nuclei with moderate to large values of $\mathcal{B}^2_2(\mathrm{HE})$, the rigid rotor picture is indeed valid.
Contrarily, spherical systems identified by their negative $\mathcal{B}^2_2(\mathrm{HE})$ value show the largest deviation from the pure rigid rotor behavior.

\subsection{Excited $0^+_2$ state in $^{12}$C, $^{16}$O and $^{20,22}$Ne}

\renewcommand{\arraystretch}{1.3} 
\begin{table}[ht!]
\centering
\caption{Rescaled quadrupole ($\mathcal{B}^2_{2}(\text{HE})$) and octupole ($\mathcal{B}^2_{3}(\text{HE})$) two-body mean-square eccentricities of the $0^+_1$ ground-state and $0^+_2$ excited state in $^{12}$C, $^{16}$O and $^{20,22}$Ne. Values of  the square of the intrinsic quadrupole and octupole deformation of the deformed HFB minimum is also indicated. 
}
\label{tab:exampleSM}
\setlength{\tabcolsep}{3pt} 
\begin{ruledtabular}
\begin{tabular}{c|c|cc|cc|cc}
\multirow{2}{*}{Nucleus} & State &  \multicolumn{2}{c}{dHFB} & \multicolumn{2}{c}{PHFB} & \multicolumn{2}{c}{PGCM} \\ 
 &  & $\beta^2_{20}$ & $\beta^2_{30}$ & $\mathcal{B}^2_2$ 
& $\mathcal{B}^2_3$ & $\mathcal{B}^2_2$  & $\mathcal{B}^2_3$  \\  \hline
$^{12}\mathrm{C}$  & $0^+_1$  & $+0.13$ & $+0.00$ & $+0.07$ & $+0.00$ & $+0.10$ & $+0.01$ \\
                   & $0^+_2$  &         &         &         &         & $+1.25$ & $-0.28$ \\  \hline
$^{16}\mathrm{O}$  & $0^+_1$  & $+0.00$ & $+0.00$ & $-0.06$ & $+0.00$ & $-0.05$ & $+0.18$ \\
                   & $0^+_2$  &         &         &         &         & $-0.05$ & $+0.10$ \\  \hline
$^{20}\mathrm{Ne}$ & $0^+_1$  & $+0.19$ & $+0.00$ & $+0.13$ & $-0.07$ & $+0.21$ & $+0.10$ \\
                   & $0^+_2$  &         &         &         &         & $+0.33$ & $+0.21$ \\  \hline
$^{22}\mathrm{Ne}$ & $0^+_1$  & $+0.18$ & $+0.00$ & $+0.13$ & $-0.07$ & $+0.18$ & $-0.01$ \\
                   & $0^+_2$  &         &         &         &         & $+0.44$ & $+0.19$ \\
\end{tabular}
\end{ruledtabular}
\end{table}

The rescaled quadrupole ($\mathcal{B}^2_{2}(\text{HE})$) and octupole ($\mathcal{B}^2_{3}(\text{HE})$) two-body mean-square eccentricities obtained from PGCM calculations of $^{12}$C, $^{16}$O and $^{20,22}$Ne were analyzed in the body of the paper for the $0^+_1$ ground-state. Table~\ref{tab:exampleSM} complements this with the data corresponding to the $0^+_2$ excited state. Starting with $^{12}$C, mean square eccentricities of the Hoyle state  differ strongly from those of the ground state, i.e., gigantic quadrupole correlations are predicted with $\mathcal{B}^2_{2}(\text{HE})=+1.25$, along with $\mathcal{B}^2_{3}(\text{HE})$ being large and negative. While the $0^+_2$ state in $^{16}$O resembles the ground state, it displays much larger $\mathcal{B}^2_{2}(\text{HE})$ and $\mathcal{B}^2_{3}(\text{HE})$ values in $^{20,22}$Ne. These results illustrate how constraining ground-state correlations via high-energy ion-ion collisions can then allow {\it ab initio} theory to predict properties of even more strongly correlated excited states.

\subsection{Kumar effective deformation parameter}

As stipulated in the main-body of the paper, the inclusion of the trivial contribution originating from the one-body part of the Kumar operator $\mathcal{Q}^{(2)}_{\ell}$ into the definition of the effective deformation parameter $\mathcal{B}^2_\ell(\mathrm{LE})$ is responsible for the fact that the latter does not correlate well with the intrinsic deformation parameter $\beta^2_{\ell 0}$ of the (quantum) rotor model, i.e., the PHFB approximation. This is at variance with the effective deformation parameter $\mathcal{B}^2_\ell(\mathrm{HE})$ that does not include such a one-body term and that correlates well with $\beta^2_{\ell 0}$. The marked difference between both effective deformation parameters demonstrate that the inclusion of the completeness relation to express $\langle \mathcal{Q}^{(2)}_{\ell} \rangle$ in terms of reduced electromagnetic transitions probabilities and the application of the RR model to infer an effective deformation parameter from it constitute two operations that do not commute. Indeed, applying the RR model after the inclusion of the completeness relation wrongly delivers  $\langle \mathcal{Q}^{(2)}_{\ell} \rangle^{\text{RR}}_{0^+} \propto \beta^2_{\ell 0}$ and thus leads to overlooking the contribution of the one-body term to $\langle \mathcal{Q}^{(2)}_{\ell} \rangle_{0^+}$.

This key difference between both effective deformation parameters is presently investigated in more detail, both analytically and numerically. For this purpose, another effective parameter $\Tilde{\mathcal{B}}^2_2(\mathrm{HE})$ is defined by \emph{wrongly} including the one-body contribution to the mean-square eccentricity according to
\begin{align}
    \label{eq: SM Tilde B2(HE) def}
    \Tilde{\mathcal{B}}^2_2(\mathrm{HE}) &\equiv \frac{4\pi}{3} \langle \delta \epsilon^2_2 \rangle \nonumber \\
    &= \frac{2\pi}{3} \frac{\langle [\mathcal{E}^{(2)}_{2}]_{00} \rangle}{\langle R^{(1)}_2 \rangle^2} \nonumber \\
    &= \frac{16\pi^2}{25} \frac{\langle \mathcal{Q}^{(2)}_{2}(p+n) \rangle}{\langle r^2 \rangle^2}\; ,
\end{align}
where $\mathcal{Q}^{(2)}_{2}(p+n)$ extends the definition of the Kumar operator $\mathcal{Q}^{(2)}_{2}$ by summing over both protons and neutrons. 

To quantify the role of the incriminated one-body term, two different ratios of effective deformation parameters
\begin{subequations}
\begin{align}
\label{eq: SM ratios}
\frac{\mathcal{B}^2_2(\mathrm{LE})}{\Tilde{\mathcal{B}}^2_2(\mathrm{HE})} &= \frac{4 \langle \mathcal{Q}^{(2)}_{2} \rangle}{\langle \mathcal{Q}^{(2)}_{2}(p+n) \rangle} \, , \\
\frac{\mathcal{B}^2_2(\mathrm{LE})}{\mathcal{B}^2_2(\mathrm{HE})} &= \frac{4 \langle \mathcal{Q}^{(2)}_{2} \rangle}{\langle \mathcal{Q}^{(2)\, \mathrm{2b}}_{2}(p+n) \rangle}  \, ,
\end{align}
\end{subequations}
are investigated. To mitigate the fact that $\mathcal{B}^2_2(\mathrm{LE})$ involves only protons while $\mathcal{B}^2_2(\mathrm{HE})$ and $\Tilde{\mathcal{B}}^2_2(\mathrm{HE})$ sum over both protons and neutrons, the analysis is presently restricted to $Z=N$ nuclei. In this case, the one-body and two-body components of the operator $\mathcal{Q}^{(2)}_{2}(p+n)$ satisfy\footnote{These identities are strictly valid if omitting the Coulomb interaction and the isospin breaking contributions to the strong interaction.}
\begin{subequations}
\begin{align}
    \mathcal{Q}^{(2)\, \mathrm{1b}}_{2}(p+n) = 2\mathcal{Q}^{(2)\, \mathrm{1b}}_{2} \; ,\\
    \mathcal{Q}^{(2)\, \mathrm{2b}}_{2}(p+n) = 4\mathcal{Q}^{(2)\, \mathrm{2b}}_{2} \; .
\end{align}
\end{subequations}
such that both ratios simplify to
\begin{subequations}
\begin{align}
\label{eq: SM ratio B2(LE) tilde B2(HE)}
\left[\frac{\mathcal{B}^2_2(\mathrm{LE})}{\Tilde{\mathcal{B}}^2_2(\mathrm{HE})}\right]_{N=Z} &= \frac{4 \langle \mathcal{Q}^{(2)\, \mathrm{1b}}_{2} \rangle + 4 \langle \mathcal{Q}^{(2)\, \mathrm{2b}}_{2} \rangle}{2 \langle \mathcal{Q}^{(2)\, \mathrm{1b}}_{2} \rangle + 4 \langle \mathcal{Q}^{(2)\, \mathrm{2b}}_{2} \rangle}\; , \\
\left[\frac{\mathcal{B}^2_2(\mathrm{LE})}{\mathcal{B}^2_2(\mathrm{HE})}\right]_{N=Z} &= \frac{4 \langle \mathcal{Q}^{(2)\, \mathrm{1b}}_{2} \rangle + 4 \langle \mathcal{Q}^{(2)\, \mathrm{2b}}_{2} \rangle}{4 \langle \mathcal{Q}^{(2)\, \mathrm{2b}}_{2} \rangle} \, .
\end{align}
\end{subequations}
Further applying the RR model eventually leads to the characteristic forms 
\begin{subequations}
\label{eq: SM ratios RR}
\begin{align}
\left[\frac{\mathcal{B}^2_2(\mathrm{LE})}{\Tilde{\mathcal{B}}^2_2(\mathrm{HE})}\right]^{\mathrm{RR}}_{N=Z} &=  1 + \frac{1}{1 + \frac{3A\beta^2_{20}}{8\pi}}  \, , \label{eq: SM ratios RR_1} \\
\left[\frac{\mathcal{B}^2_2(\mathrm{LE})}{\mathcal{B}^2_2(\mathrm{HE})}\right]^{\mathrm{RR}}_{N=Z} &=  1 +\frac{8\pi}{3A\beta^2_{20}} \, . \label{eq: SM ratios RR_2}
\end{align}
\end{subequations}

\begin{figure}[h]
    \centering
    \includegraphics[width=\columnwidth]{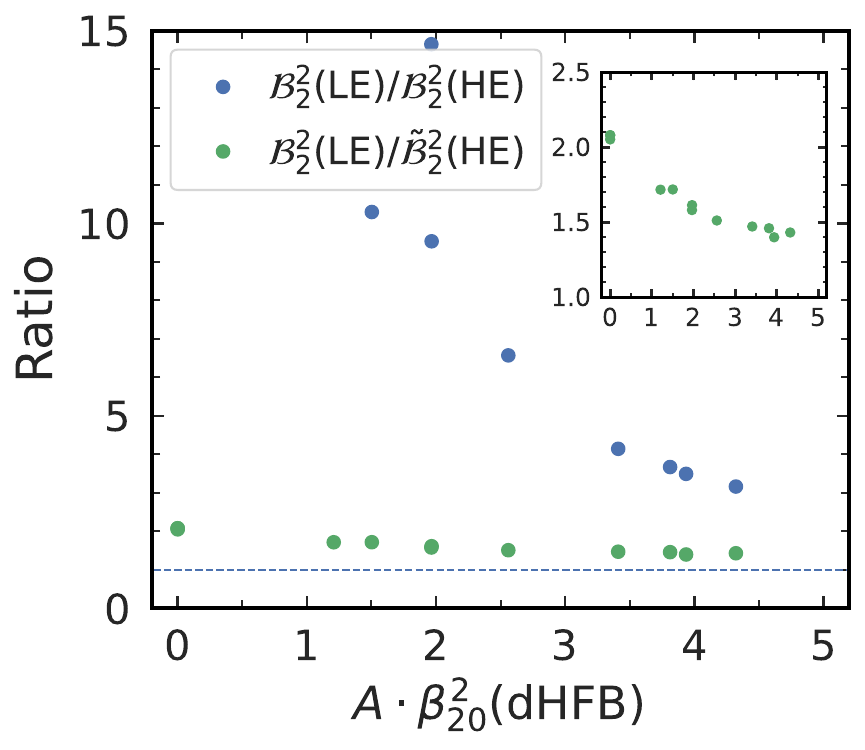}
    \caption{Ratios $\mathcal{B}^2_2(\mathrm{LE})/\mathcal{B}^2_2(\mathrm{HE})$ and $\mathcal{B}^2_2(\mathrm{LE})/\Tilde{\mathcal{B}}^2_2(\mathrm{HE})$ from PHFB calculations of $N=Z$ nuclei between carbon and nickel as a function of $A \cdot \beta^2_{20}(\mathrm{dHFB})$. The inset zooms over the ratios ranging between $1.0$ and $2.5$.}
    \label{fig: SM ratio B2(LE) tildeB2(HE)}
\end{figure}

Both ratios are displayed in Fig.~\ref{fig: SM ratio B2(LE) tildeB2(HE)} for PHFB results of $N=Z$ nuclei between carbon and nickel as a function of the product $A \cdot \beta^2_{20}(\text{dHFB})$. The observed trend of both ratios is fully consistent with the RR analytical result obtained in Eq.~\eqref{eq: SM ratios RR}.

Overall, the $\mathcal{B}^2_2(\mathrm{LE})/\Tilde{\mathcal{B}}^2_2(\mathrm{HE})$ is much closer to $1$ than $\mathcal{B}^2_2(\mathrm{LE})/\mathcal{B}^2_2(\mathrm{HE})$, which indeed demonstrates that the inappropriate inclusion of the one-body term into $\mathcal{B}^2_2(\mathrm{LE})$ generates the difference from $\mathcal{B}^2_2(\mathrm{HE})$. Let us further analyze the behavior of both ratios.

For light nuclei displaying small intrinsic deformation, the dominance of the one-body term combined with the different isospin character of both operators at play result into $\mathcal{B}^2_2(\mathrm{LE})/\Tilde{\mathcal{B}}^2_2(\mathrm{HE}) \sim 2 $ (Eq.~\eqref{eq: SM ratios RR_1})\footnote{The negative contribution of the exchange term in spherical systems generates in fact values that are slightly larger than 2.}. As predicted by the RR model (Eq.~\eqref{eq: SM ratios RR_2}), the ratio decreases towards 1 as the product $A \cdot \beta^2_{20}(\mathrm{dHFB})$ increases due to the dominance of the two-body contribution to both operators in this limit. 
However, in the set of nuclei under study there is no system for which the two-body term is sufficiently large to effectively approach 1; i.e., for the largest value of $A \cdot \beta^2_{20}(\mathrm{dHFB})$ presently reached, the ratio remains half way between $2$ and $1$ as the inset shows.

More importantly, the relevant ratio $\mathcal{B}^2_2(\mathrm{LE})/\mathcal{B}^2_2(\mathrm{HE})$ is much larger than $1$ for most of the nuclei under study, thus confirming the inoperative character of the effective deformation parameter $\mathcal{B}^2_2(\mathrm{LE})$ extracted from the Kumar operator. Whenever $\beta^2_{20}(\mathrm{dHFB})$ is small, the second term in Eq.~\eqref{eq: SM ratios RR_2} is further enhanced due to the presently neglected exchange term that effectively reduces the denominator\footnote{In fact, $\mathcal{B}^2_2(\mathrm{HE})$ is eventually negative in the limit $\beta^2_{20}(\mathrm{HFB})\approx 0$. The corresponding points are not depicted in Fig.~\ref{fig: SM ratio B2(LE) tildeB2(HE)}.}. As the product $A \cdot \beta^2_{20}(\mathrm{dHFB})$ increases, the ratio $\mathcal{B}^2_2(\mathrm{LE})/\mathcal{B}^2_2(\mathrm{HE})$ decreases as $(A \cdot \beta^2_{20}(\mathrm{dHFB}))^{-1}$ to eventually reach $1$ in well-deformed heavy nuclei. For the nuclei under study whose maximum mass is $A=80$, this ratio is however never smaller than $3$.


\bibliographystyle{apsrev4-1}
\bibliography{biblio}

\end{document}